\documentclass[journal]{IEEEtran}
\usepackage{mathrsfs}
\usepackage{amsfonts}
\usepackage{ifpdf}
\usepackage{cite}
\ifCLASSINFOpdf
\usepackage[pdftex]{graphicx}
\else \fi
\usepackage[cmex10]{amsmath}
\usepackage{array}
\usepackage{mdwmath}
\usepackage{mdwtab}
\usepackage{eqparbox}
\usepackage[tight,footnotesize]{subfigure}
\usepackage{algorithmic}
\usepackage{algorithm}
\usepackage{amsmath}
\usepackage{epsfig}
\usepackage{ae}
\usepackage{amssymb}
\usepackage{times}
\usepackage{amsmath}   
\usepackage{booktabs}  
\usepackage{threeparttable} 
\usepackage{multirow}       
\usepackage{xcolor}
\usepackage{graphicx}
\usepackage{subfigure}

\newtheorem{theorem}{\textbf{Theorem}}
\newtheorem{lemma}{\textbf{Lemma}}

\newtheorem{remark}{\textbf{Remark}}
\newtheorem{proposition}{\textbf{Proposition}}

\begin{document}
%
\title{Optimizing Information Freshness in Computing enabled IoT Networks}

\author{Chao Xu, {\em Member, IEEE}, Howard H. Yang, {\em Member, IEEE}, Xijun Wang, {\em Member, IEEE}, and \\ Tony Q. S. Quek, {\em Fellow, IEEE}
\thanks{

This paper is supported by National Natural Science Foundation of China (61701372), Talents Special Foundation of Northwest A\&F University (Z111021801), Key Research and Development Program of Shaanxi (2019ZDLNY07-02-01), Fundamental Research Funds for the Central Universities of China (SYSU: 19lgpy79), and Research Fund of the Key Laboratory of Wireless Sensor Network \& Communication (Shanghai Institute of Microsystem and Information Technology, Chinese Academy of Sciences) under grant 20190912. This paper was presented in part in the Proc. of IEEE WCNC 2019\cite{PAoI_WCNC_2019}. (\emph{Corresponding author: Xijun Wang})

C. Xu is with School of Information Engineering, Northwest A\&F University, Yangling, Shaanxi, China (e-mail: cxu@nwafu.edu.cn). C. Xu is also with Key Laboratory of Agricultural Internet of Things, Ministry of Agriculture and Rural Affairs, Yangling, Shaanxi, China, and Shaanxi Key Laboratory of Agricultural Information Perception and Intelligent Service, Yangling, Shaanxi, China.

H. H. Yang and T. Q. S. Quek are with the Information System Technology and Design Pillar, Singapore University of Technology and Design, Singapore 487372 (e-mail: howard\_yang@sutd.edu.sg; tonyquek@sutd.edu.sg).

X. Wang is with School of Electronics and Communication Engineering, Sun Yat-sen University, Guangzhou, China (e-mail: \mbox{wangxijun@mail.sysu.edu.cn}). Xijun Wang is also with Key Laboratory of Wireless Sensor Network \& Communication, Shanghai Institute of Microsystem and Information Technology, Chinese Academy of Sciences, 865 Changning Road, Shanghai 200050 China.

Copyright (c) 2019 IEEE. Personal use of this material is permitted. However, permission to use this material for any other purposes must be obtained from the IEEE by sending a request to pubs-permissions@ieee.org.
}
} \maketitle
\begin{abstract}
Internet of Things (IoT) has emerged as one of the key features of the next generation wireless networks, where timely delivery of status update packets is essential for many real-time IoT applications. To provide users with context-aware services and lighten the transmission burden, the raw data usually needs to be preprocessed before being transmitted to the destination. However, the effect of computing on the overall information freshness is not well understood. In this work, we first develop an analytical framework to investigate the information freshness, in terms of peak age of information (PAoI), of a computing enabled IoT system with multiple sensors. Specifically, we model the procedure of computing and transmission as a tandem queue, and derive the analytical expressions of the average PAoI for different sensors. Based on the theoretical results, we formulate a min-max optimization problem to minimize the maximum average PAoI of different sensors. We further design a derivative-free algorithm to find the optimal updating frequency, with which the complexity for checking the convexity of the formulated problem or obtaining the derivatives of the object function can be largely reduced. The accuracy of our analysis and effectiveness of the proposed algorithm are verified with extensive simulation results.
\end{abstract}

\section{Introduction}\label{sec:introduction}
Being one of the key technologies of the next generation (5G) wireless networks, Internet of Things (IoT) has attracted significant attentions from both academia and industry alike in recent years. In particular, IoT aims at enabling the ubiquitous connectivity among billions of things, ranging from tiny, resource-constrained sensors to more powerful smartphones and networked vehicles \cite{IoT_5G_2016,IoT_5G_2017,Fu_SN_2018}. With the help of IoT, devices can sense and even interact with the physical surrounding environment, thereby providing us with many valuable and remarkable context-aware real time applications at an efficient cost, such as automatic control of electric appliance \cite{Conference_Transit_Data}, intelligent transportation network \cite{Traffic_IoT},  and event monitoring and predication for health safety \cite{Data_Mining_IoT}.
For these applications, the staleness of obtained information at destinations inevitably deteriorates the accuracy and reliability of derived decisions, and even compromises in safety and security. In order to quantify the information freshness, age of information (AoI) \cite{AoI_Org_2012} and peak age of information (PAoI) \cite{PAoI_2014_ISIT} have been recently introduced. Particularly, AoI measures the time elapsed since the latest received update packet was generated, while PAoI provides information about the maximum value of AoI for each update and captures the extent to which the update information is stale. Unlike many conventional metrics, e.g., delay or throughput\cite{Dealy_Throughput_2004,Distributed_Subchannel_allo_CXu,Liu_VN_2019}, AoI and PAoI are affected not only by the transmission delay but also by the update generation rate, and hence they are more essential and comprehensive for information freshness evaluation\cite{AoI_Survey_2017}.

In conventional IoT networks, due to the limited communication resource, a significant delay may occur during the packet transmission phase, which  largely deteriorates the information freshness at the receiver side.
As discussed in a line of existing work \cite{IoT_Context_Aware_2014,IoT_Context_Aware_2015,IoT_Context_Aware_2018}, to lighten the transmission burden and provide the end users with better context-aware services in IoT networks, it is preferable to first process the collected raw data with the edge/fog computing technique \cite{MEC_IoT_White_Paper_2015,MEC_IoT_2017,MEC_IoT_2018}, and then transmit the resultant packet, which has a large reduction in size, to the actuator or monitor.
However, the effect of such a data preprocessing procedure on the information freshness has not been fully understood.

In this paper, we consider a computing enabled IoT network, which consists of multiple sensors, a data aggregator, and a destination. The aggregator first preprocesses the status updates generated from sensors with different priorities and then forwards the processed data to the destination via a wireless channel according to the first-come-first-serve (FCFS) discipline. By modeling the system as a tandem queue, we analytically derive the expressions of the average PAoI for updates from different sensors, accounting for the joint effect of data preprocessing and transmission. Furthermore, we develop a Generating set search based Average PAoI minimization (GAP) algorithm to optimally control the generation rate of updates from different sensors to achieve the best information freshness. The accuracy of our analysis and the effectiveness of our proposed GAP algorithm are verified with extensive simulations. The main contributions of this work can be summarized as follows.
\begin{itemize} \setcounter{enumi}{0}
\item We establish a mathematical framework to model the joint effect from data preprocessing and transmission on the information freshness of an IoT network. Our framework is general and captures many key features in IoT networks, including the prioritized data processing, queueing, and wireless channel fading. We respectively derive a closed-form expression and an information theoretic approximation of the expectation of waiting time for the data processing queue and transmission queue. Based on these, we obtain the analytical expressions of the average PAoI for packets from different sensors. The accuracy of our analysis is verified via simulations, which shows a good match between the simulation and theoretical results.
\item We develop a derivative-free GAP algorithm to search for the solution of the formulated min-max programming, which minimizes the maximum average PAoI for updates from different sensors. Particularly, with GAP, the problem can be solved by getting around of the difficulty of checking the convexity of the formulated problem or resorting to the derivatives of the object function. Due to this features, our proposed algorithm is still available when other penalty functions (instead of the maximum of average PAoI) are incorporated. Besides, the global convergence of GAP is also presented. The convergence rate of GAP is acceptable and does not exponentially increase with number of sensors. Hence, it still works when the number of sensors becomes large. Moreover, compared with the baseline strategies, the achieved maximum average PAoI can be effectively reduced by implementing our proposed GAP algorithm.
\end{itemize}

The outline of this paper is as follows. In Section II, a brief survey of related work is presented. The description of system model and mathematical definitions of AoI and PAoI are given in Section III. In Section IV, we analyze the average PAoI for updates from different sensors and verify the accuracy of our analysis via simulations. In Section V, based on the obtained analytical results, we formulate a min-max programming to minimize the achieved maximum average PAoI of sensors, develop the GAP algorithm to solve it, and conduct simulations to validate the effectiveness of this algorithm. Finally, conclusions are drawn in Section VI.

\section{Related Work}

Ever since the concept of AoI was introduced,
a variety of researches have been carried out to understand and/or optimize the information freshness of the delivered update packets in single sensor systems\cite{AoI_Org_2012,LIFO_AoI,PAoI_2014_ISIT,PAoI_TIT_2016,AoI_HARQ,IF_GG1_Distributation,Age_G/G/1/1_2018,Dataa_Age_TIT,Estimation_AoI_Wiener,Estimation_AoI_Markov}. Authors in \cite{AoI_Org_2012} considered the system where a sensor generated and transmitted update packets to its destination with the FCFS principle and derived the expression of average AoI by resorting to a queueing theoretic approach. Then, AoI of the last-come-first-served (LCFS) queueing based system was further studied in \cite{LIFO_AoI}, and it demonstrated that the AoI was improved compared with the FCFS based system. In \cite{PAoI_2014_ISIT} and its journal version \cite{PAoI_TIT_2016}, the effects of packet preemption on both AoI and PAoI were respectively analyzed by considering three distinct preemption policies. Authors in \cite{AoI_HARQ} further considered a symbol erasure transmission channel and studied the effect of packet preemption on the average AoI when adopting two hybrid ARQ protocols. For IoT networks with arbitrary distributions of the update
inter-arrival time and service time, the relation among the distributions of the AoI, PAoI and system delay was derived in \cite{IF_GG1_Distributation}, while the effect of packet preemption on the AoI was investigated in \cite{Age_G/G/1/1_2018}. Focusing on the one hop transmission from a data source to a destination, authors in \cite{Dataa_Age_TIT} introduced a general penalty function to characterize the effect of AoI, and developed efficient algorithms to find the optimal update policy for minimizing the average penalty among all causal update policies. In \cite{Estimation_AoI_Wiener} and \cite{Estimation_AoI_Markov} the effects of sampling strategies on the tradeoff between the achieved AoI and estimation accuracy for remote estimation problems was addressed, where the environment related state was assumed to be generated from a discrete Markov process and Wiener process, respectively.

Apart from the point-to-point scenario \cite{AoI_Org_2012,LIFO_AoI,PAoI_2014_ISIT,PAoI_TIT_2016,AoI_HARQ,IF_GG1_Distributation,Age_G/G/1/1_2018,Dataa_Age_TIT,Estimation_AoI_Wiener,Estimation_AoI_Markov}, a line of recent studies turned their attention to addressing the information freshness related issues in IoT networks with multiple sensors \cite{Multi_Source_2015_ICC,Multi_Source_ISIT,Multi_Source_ARXIV,Multi_SOurce_Preemptive_2018,Max_Min_PAoI,Multi_Source_Multi_D_SIngle_T,Multi_Source_Multi_D_Multi_T}. In particular, authors in \cite{Multi_Source_2015_ICC} considered that one transmitter sent status update packets generated from multiple sensors to the destination, and analyzed the average AoI for updates allowing the latest arrival to overwrite the previous queued ones. Focusing on the PAoI metric, work \cite{Multi_Source_ISIT} analyzed the system performance by considering a general service time distribution, and tried to optimize the update arrival rates to minimize its defined PAoI-related system cost. In \cite{Multi_Source_ARXIV} the AoI was thoroughly studied for systems with three different serving policies, i.e., FCFS, LCFS with preemption in service, and LCFS with preemption only in waiting. Furthermore, in \cite{Multi_SOurce_Preemptive_2018} preemption of pacckets were allowed for FCFS based transmission when the transmitter was busy, and expressions of both the average AoI and PAoI were derived. Although the priority issue was not specifically addressed, the authors in \cite{Multi_SOurce_Preemptive_2018} deduced that updates from one sensor can be prioritized from the age point of view by increasing their generation rate.
Authors in \cite{Max_Min_PAoI} considered the interactions among transmission links and proposed link scheduling algorithms to minimize the maximum PAoI of update packets from different sensors. For multi-sensor multi-destination networks, the AoI oriented optimal scheduling policy was studied in \cite{Multi_Source_Multi_D_SIngle_T} and \cite{Multi_Source_Multi_D_Multi_T}, which considered the scenario with one transmitter and multiple transmitters, respectively.

As mentioned above, valuable performance analysis on information freshness and efficient control strategies for minimizing the system AoI or PAoI in various IoT networks with multiple sensors have been presented in the literature \cite{Multi_Source_2015_ICC,Multi_Source_ISIT,Multi_Source_ARXIV,Multi_SOurce_Preemptive_2018,Max_Min_PAoI,Multi_Source_Multi_D_SIngle_T,Multi_Source_Multi_D_Multi_T}. However, they commonly treated the data aggregator purely as a transmitter and thus do not apply to the case, where update packets would be preprocessed (e.g., data compression and aggregation) to reduce the redundancy or even extract the ``intrinsic content'' from the collected raw data, before any transmission procedure begins.
Actually, the joint operation of data preprocessing and transmission has been regarded as a promising solution for providing better context-aware services to users and meanwhile, overcoming the resource limitations on transmission capacity inherent in traditional IoT networks\cite{IoT_Context_Aware_2014,IoT_Context_Aware_2015,IoT_Context_Aware_2018}. As such, it calls for additional efforts to study and optimize the information freshness when the data preprocessing and transmission are successively conducted.
The most related work to this topic comes from \cite{PAoI_Camera_Networks}, which studied the wireless camera networks consisting of multiple sensors and fog nodes, and proposed a modular optimization algorithm to minimize the achieved maximum PAoI by optimally assigning processing nodes and scheduling transmission links. However, the joint effect of the processing procedure (e.g., processing policy and time) and update arrival rates on information freshness has not been investigated in \cite{PAoI_Camera_Networks} nor, to the best of our knowledge, in other existing researches.

We note that there are some available work recently focusing on studying and/or optimizing AoI and/or PAoI for status updates in multi-hop IoT networks, e.g., \cite{Multi_Hop_LGFS,Multi_Hop_No_Queue_2018,Two_Hopes_EH_AoI,Graph_AoI_2017,Graph_AoI_PAoI_Bound}, which are also relevant to our work. Particularly, authors in \cite{Multi_Hop_LGFS} considered a multi-hop networks with an external source, and proved that, among all causal policies, the preemptive Last Generated First Served (LGFS) policy and non-preemptive LGFS policy minimized the age processes at all nodes for the exponentially distributed and generally distributed packet transmission times, respectively. Authors in \cite{Multi_Hop_No_Queue_2018} focused on a line network with one sensor, one destination, and multiple relay nodes, and studied the effect of preemption on the average AoI at each node. The energy and data causality constraints in a two-hop network were considered in \cite{Two_Hopes_EH_AoI}, and the optimal scheduling policy was proposed to minimize the total AoI of a session. Considering multi-hop networks and utilizing graph theory, the optimal scheduling policies for the networks with and without pre-defined source/destination pairs were investigated in \cite{Graph_AoI_2017} and \cite{Graph_AoI_PAoI_Bound}, respectively. However, the aggregator considered in our work plays different roles to those relay nodes studied in \cite{Multi_Hop_LGFS,Multi_Hop_No_Queue_2018,Two_Hopes_EH_AoI,Graph_AoI_2017,Graph_AoI_PAoI_Bound}. Particularly, the aggregator not only forwards the packets, but also regenerates the packets with different sizes, which would alter the service time of the second queue. This makes the interaction between the aggregator and transmitter more complicated than that between relay nodes purely for data retransmission. In this light, our concerned problem, formulated mathematical model, derived analysis results and proposed optimizing algorithm are all different from those presented in existing researches\cite{Multi_Hop_LGFS,Graph_AoI_2017,Two_Hopes_EH_AoI,Multi_Hop_No_Queue_2018,Graph_AoI_PAoI_Bound}.

\section{System Model}

\begin{figure} [!t]
\centering \includegraphics[width=3.0in]{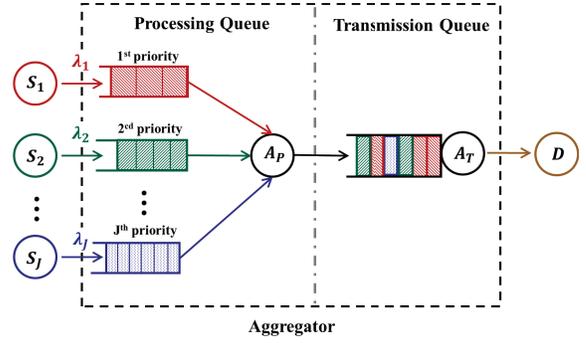}
\centering \caption{Illustration of the tandem queueing model for the considered IoT network.} \label{Fig:Fig1}
\end{figure}

\subsection{Network Model}
We consider an IoT system which consists of $J$ sensors, denoted by $\mathcal{S} = \{S_1, S_2, \cdots, S_J\}$, a data aggregator that is able to perform data processing as well as transmission, and a destination node, as depicted in Fig.~\ref{Fig:Fig1}. Each sensor keeps collecting information from the ambient environment and periodically updates the status to the aggregator, whereas the update packets from sensor $S_j$ arrive at the aggregator according to an independent Poisson process with parameter $\lambda_j, \forall j \in \mathcal J = \{1, 2, \cdots, J\}$.
Upon receiving the status updates, the aggregator  preprocesses the data packets with different priorities and then forwards them to the destination node.
Without loss of generality, we assume the data from $S_i$ has a higher priority than that from $S_j$ if $i < j$. In this regard, a generic update packet can only be processed if, in front of it, there is no packet with a higher or equal priority being or waiting to be processed.

For an incoming update packet with the $j$-th priority, we denote $C_j$ and $\tilde{C}_j$ ($\tilde{C}_j<C_j$) as the size before and after data processing, respectively, and $\tau_j\frac{C_j-\tilde{C}_j}{r}$ the corresponding processing time. Here, $r$ is the CPU's computational speed of the aggregator with the units CPU cycles per second, while $\tau_j$ is a scaling parameter depends on the specific operation made on the packet and with the units CPU cycles per bit. For instance, data mining may be more complicated than data compression and would be endowed with a larger $\tau$. We note that the similar computation model has been widely used for data processors as shown in \cite{Survey_Com_Time_2017} and references therein. As such, the preprocessing subsystem is formulated as a priority M/G/1 queue where the size of buffer is infinite.\footnote{It should be noted that the following analysis also holds when we consider another processing model mapping each $(C_j, \tilde{C}_j)$ to a positive real number (i.e., the processing time) since a priority M/G/1 queue can also be formulated in that scenario.}

After being preprocessed, each update packet will be pushed into an infinite-size queue at the transmitter according to the FCFS discipline. We term this buffer the transmission queue. The transmitter sends each packet with a constant power $p_A$ through a channel with bandwidth of $B$ Hz. The channel is subjected to a small scale Rayleigh fading with unit mean and a large scale path loss that follows power law, with path loss exponent $\alpha > 2$. Both the processing queue and transmission queue are considered to be non-preemptive.

\begin{figure}[!t]
\centering \includegraphics[width=3.0in]{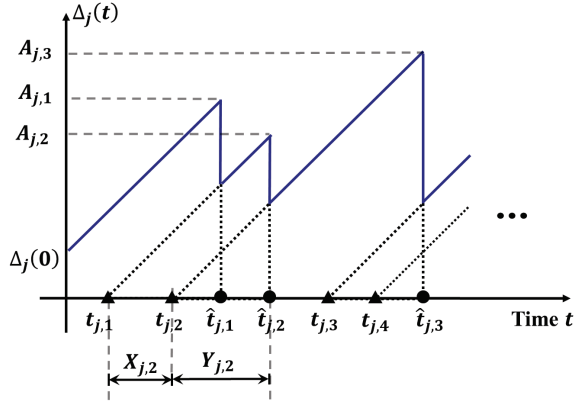}
\centering \caption{Example of the AoI evolution process for sensor $S_j$ at the destination node. The time instant of packet arrival at the aggregator and the destination node are marked as $\blacktriangle$ and $\bullet$, respectively.} \label{Fig:Age_variation_Example}
\end{figure}

\subsection{Age of Information}
We denote $t_{j,n}$ the time instant when the $n$-th packet from $S_j$ arriving at the aggregator,\footnote{Similar as previous studies \cite{Multi_Source_ISIT,Multi_Source_2015_ICC,Multi_SOurce_Preemptive_2018,Multi_Source_ARXIV}, we
consider the time spent on the transmission from sensors to the aggregator negligible since they are generally integrated as a whole system and connected via high speed wired links.} and denote with $\hat{t}_{j,n}$ the time instant that this packet arrived at the destination node. The AoI of sensor $S_j$ is defined as ${\Delta _j}( t ) = t - {u_j}( t )$, where ${u_j}( t )$ is the generation time of the most recently received packet from $S_j$ until time instant $t$ \cite{Multi_Source_ISIT,Multi_Source_2015_ICC,Multi_SOurce_Preemptive_2018,Multi_Source_ARXIV}.
An example of the AoI evolution process ${\Delta _j}(t)$ for the $j$-th sensor is illustrated in Fig. \ref{Fig:Age_variation_Example}.
It can be seen that after one packet arrived at the destination node, the AoI increases linearly in time until a new data packet is received. In other words, the $n$-th peak value of ${\Delta _j}(t)$ is achieved just before the $n$-th update packet arrives at the destination node, which is defined as the PAoI and denoted by $A_{j,n}$ as shown in Fig. \ref{Fig:Age_variation_Example}.
Formally, the PAoI evolves as follows
\begin{align} \label{Eq:Peak_Age_Information_Instance}
A_{j,n}=
\begin{cases}
{\Delta _j}\left( 0 \right)+\hat{t}_{j,n}, &n=1 \\
X_{j,n}+Y_{j,n}, & n >1
\end{cases}
\end{align}
where ${\Delta _j}( 0 )$ denotes the initial age of the last received data at the start time, $X_{j,n}$ represents the time interval between $t_{j,n}$ and $t_{j,n-1}$, and $Y_{j,n}$ represents the time interval between $\hat{t}_{j,n}$ and $t_{j,n}$, i.e., $X_{j,n}= t_{j,n}-t_{j,n-1}$ and $Y_{j,n}=\hat{t}_{j,n}-t_{j,n}$.
It is worth noting that while the inter-arrival time $X_{j,n}$ only relates to the sensor $S_j$, the system time  $Y_{j,n}$ is determined by many factors, including the packet arrival processes from $S_j$ and $\mathcal S_{-j}=\{S_1, S_2, \cdots, S_{j-1}, S_{j+1}, \cdots, S_J\}$, and the preprocessing and transmission processes in the aggregator. As such, we can write $Y_{j,n}$ as the sum of the time that the packet $n$ spent in the preprocessing stage $Y_{j,n}^P$ and that in the transmission stage $Y_{j,n}^T$, i.e., $Y_{j,n}=Y_{j,n}^P+Y_{j,n}^T$.

The central thrust of this work is to design a scheme that ensures the received packets contain the most fresh information.
To achieve this goal, we first derive an analysis and then conduct optimization on the achieved average PAoI for updates. In the following, we provide detailed analyses for the joint effects of the preprocessing and transmission procedures on the achieved average PAoI for updates from different sensors in Section \ref{Sec:PAoI_Analysis}. Based on the analytical results, to minimize the achieved maximum average PAoI by controlling the update generation rates of individual sensors, we formulate a min-max programming and devise a derivative-free algorithm to solve it in Section \ref{Sec:PAoI_Optimization}. Some important notations used in this paper are summarized in Table \ref{Tab:Notations}.

\begin{table}[!t]
\renewcommand{\arraystretch}{1.3}
\caption{Description of important notations.}
\label{Tab:Notations}
\centering
\begin{tabular}{| l | l | }
\hline
Notation                                                & Description     \\
\hline
$J$                                                     & Number of sensors     \\
\hline
$\mathcal S$                                            & Set of sensors    \\
\hline
$\lambda_j$                                             & Arrival rate of packets from sensor $S_j$     \\
\hline
$C_j$                                                   & Size of the original packet for sensor $S_j$    \\
\hline
$\tilde{C}_j$                                           & Size of the processed packet for sensor $S_j$  \\
\hline
\multirow{2}{*}{$\frac{r}{\tau_j}$}                     & {Equivalent data processing rate for packets from } \\
 {}                                                     & {sensor $S_j$} \\
\hline
$p_A$                                                   & Transmit power of the aggregator \\
\hline
$B$                                                     & Transmission bandwidth \\
\hline
$d$                                                     & Distance between the aggregator and destination node\\
\hline
$\alpha$                                                & Path loss exponent \\
\hline
${\sigma ^2}$                                           & AWGN power at the destination node\\
\hline
$\Delta_j(t)$                                           & AoI for sensor $S_j$ at time $t$  \\
\hline
\multirow{2}{*}{$A_{j,n}$}                              & {PAoI associated with the $n$-th packet arriving at } \\
 {}                                                     & {the destination from sensor $S_j$} \\
\hline
$A_j$                                                   & Average PAoI for packets from sensor $S_j$  \\
\hline
$\mathbb E \big[ {{X_j}} \big]$                         & Expected inter-arrival time of packets from sensor $S_j$  \\
\hline
$\mathbb E [ {Z_j^P} ]$                                 & Expected processing time for packets from sensor $S_j$ \\
\hline
$\mathbb E \big[ {Z_j^T} \big]$                         & Expected transmission time for packets from sensor $S_j$ \\
\hline
\multirow{2}{*}{$\mathbb E \big[ W_j^P\big]$}           & {Expected waiting time in the data processing queue for} \\
 {}                                                     & {packets from sensor $S_j$}  \\
\hline
\multirow{2}{*}{$\mathbb E \big[ {W_j^T}\big] $}        & {Expected waiting time in the data transmission queue for} \\
 {}                                                     & {packets from sensor $S_j$}  \\
\hline
$P_{A_P,B}$                                             & Busy probability of the data processor  \\
\hline
\multirow{2}{*}{$\mu_j$}                                & {Ratio of the expected waiting time for packets from sensor} \\
 {}                                                     & {$S_j$ to that from sensor $S_1$ in the transmission queue}  \\
\hline
${\mathbf {\Lambda}}$                                   &  Vector of update arrival rates from sensors \\
\hline
$\widetilde{A}\left({\mathbf {\Lambda}}\right)$         &  Achieved maximum average PAoI associated with ${\mathbf {\Lambda}}$ \\
\hline
$\mathcal I\left( {{\bf{\Lambda }},\varepsilon }\right)$      & Set of indexes of $\varepsilon$-binding constraints associated with ${\mathbf {\Lambda}}$\\
\hline
$\mathcal N({{\bf{\Lambda }},\varepsilon }) $           & Cone generated by the set of vectors in $\mathcal I\left( {{\bf{\Lambda }},\varepsilon }\right) $ \\
\hline
$\mathcal T({{\bf{\Lambda }},\varepsilon } )$           & $\varepsilon$-tangent cone for the polar of the cone $\mathcal N({{\bf{\Lambda }},\varepsilon }) $ \\
\hline
$\mathcal S_{\mathcal T( {{\bf{\Lambda } },\varepsilon } )}$      & Set of candidate searching directions in $\mathcal T({{\bf{\Lambda }},\varepsilon } )$\\
\hline
${\Phi _{\mathbf s_l} }$                                & Adopted searching step-size along the direction $\mathbf s_l$\\
\hline
\end{tabular}
\end{table}

\section{Average Peak Age of Information} \label{Sec:PAoI_Analysis}
In this section, we analyze the average PAoI for different sensors, which facilitates the subsequential optimizations. To start with, the following lemma presents a general form of the average PAoI for each sensor.
\begin{lemma}
Assuming that the whole queueing system is ergodic
, we can express the average PAoI attained for sensor $S_j$ as
\begin{align}\label{Eq:General_PAoI}
{A_j} = \frac{1}{{{\lambda _j}}} + \tau_j\frac{C_j-\tilde{C}_j}{r} + \mathbb E \big[ {W_j^P} \big] + \mathbb E \big[ {Z_j^T} \big]  + \mathbb E \big[ {W_j^T} \big] ,
\end{align}
where $\mathbb E[W_j^P]$ and $\mathbb E[{W_j^T}] $ respectively represent the expected time spent in the preprocessing queue and the transmission queue for an arbitrary update packet generated from the sensor $S_j$, and  $\mathbb E[{Z_j^T}]$ denotes the expected transmission time.
\end{lemma}
\begin{IEEEproof}
By ergodicity, the average PAoI for sensor $S_j$ can be calculated as
\begin{align} \label{Eq:Average_PAOI}
{A _j} &= \mathop {\lim }\limits_{t \to \infty } \frac{1}{{{N_j}\left( t \right)}}  \left( {{\Delta _j}\left( 0 \right) + \hat{t}_{j,1}} + \sum\limits_{n = 2}^{{N_j}\left( t \right)} {\left( {{X_{j,n}} + {Y_{j,n}}} \right)} \right) \nonumber \\
&\mathop  = \limits^{\left( a \right)} \mathbb E \left[ {{X_j} + {Y_j}} \right] = \mathbb E \big[ {{X_j}} \big] + \mathbb E \big[ {Y_j^P} \big] + \mathbb E \big[ {Y_j^T} \big],
\end{align}
where ($a$) follows from the fact that the effect of the sum ${{\Delta _j}(0) + \hat{t}_{j,1}}$ vanishes as $t$ goes to infinity. {${{N_j}( t )}$ denotes the number of update packets until time instant $t$}, and $X_j$, $Y_j^P$ and $Y_j^T$ are random variables respectively denoting the inter-arrival time, system time spent in the preprocessing queue and that in the transmission queue of an arbitrary update packet.

Recalling that the packet arrival from sensor $S_j$ follows exponential distribution with parameter $\lambda_j$, we thus have $\mathbb E\big[ {{X_j}} \big]={1}/{\lambda_j}$. Moreover, for each packet, the system time spent in the aggregator consists of the queueing time and serving time. Hence, Eq. (\ref{Eq:Average_PAOI}) can be written as
\begin{align} \label{Eq:Average_PAOI_Detail}
{A _j} &= \frac{1}{{{\lambda _j}}}+ \mathbb E \big[ {Z_j^P} \big] + \mathbb E \big[ {W_j^P} \big] + \mathbb E \big[ {Z_j^T} \big] + \mathbb E \big[ {W_j^T} \big]\\ \nonumber
&= \frac{1}{{{\lambda _j}}} + \tau_j\frac{C_j-\tilde{C}_j}{r} + \mathbb E \big[ {W_j^P} \big] + \mathbb E \big[ {Z_j^T} \big] + \mathbb E \big[ {W_j^T} \big],
\end{align}
where $\mathbb E [ {Z_j^P} ]$ is the average time spent for data preprocessing and is given as $\mathbb E [ {Z_j^P} ] =\tau_j {(C_j-\tilde{C}_j)}/{r}$. Besides, for an arbitrary update packet, $\mathbb E[W_j^P]$, $\mathbb E \big[ {Z_j^T} \big]$ and $\mathbb E[{W_j^T}] $ denote the expected waiting time in the preprocessing queue, the expected transmission time, and the expected waiting time in the preprocessing queue, respectively.
\end{IEEEproof}

In the following, we detail the analysis to each individual elements in \eqref{Eq:General_PAoI}, i.e., $\mathbb E [ {W_j^P} ]$, $\mathbb E [ {Z_j^T} ]$, and $\mathbb E [ {W_j^T} ]$.

\subsection{Calculation of  $\mathbb E \big[ {W_j^P} \big] $}
Due to prioritized processing, a newly arrived packet with priority $j$ has to wait till the completion of data processing for the following packets:
\begin{enumerate} []
\item The packet that is currently occupying the processor.
\item The packets with priorities from 1 to $j$ in the processing queue when the packet arrives.
\item The packets with priorities from 1 to $j-1$ that arrive while the typical packet is waiting for its service.
\end{enumerate}

We denote by $P_{A_P,B}$ the probability that the processor is busy. Using the Little's law \cite{Queueing_Performance}, we have the following equation 
\begin{align} \label{Eq:Busy_Probability_AP}
P_{A_P,B}  = \sum\limits_{j = 1}^{J}  {{\lambda _j} \mathbb E \big[ {Z_j^P} \big] }  = \sum\limits_{j = 1}^{J}  {{\lambda _j}{\tau _j}\frac{{{C_j} - {{\tilde C}_j}}}{r}}.
\end{align}
Based on the above analysis and definitions, we are now ready to derive $\mathbb E [ {W_j^P} ]$.

\begin{theorem} \label{Theorem:Exp_Waiting_Time_PQ}
For packets from sensor $S_j$, the expected waiting time in the processing queue is given by
\begin{align} \label{Eq:Remaining Processing_Time_Fin}
\mathbb E \big[ {W_j^P} \big] =
\frac{{\sum_{j = 1}^J {{{{{ {{\rho _j}} }^2}}}/{{{\lambda _j}}}} }}{{2\left( {1 - \sum_{i = 1}^j {{\rho _i}} } \right) \! \left( {1 - \chi_{ \{ j > 1 \} } \sum_{i = 1}^{j - 1} {{\rho _i}} } \right)}},
\end{align}
where $\rho _j={{\lambda _j} \mathbb E [{Z_j^P}] } $ denotes the load contributed by the data packets from sensor $S_j$, and $\chi_{ \{ \cdot\} }$ is the indicator function, i.e., $\chi_{ \{ j >1 \} } = 1$ if $j >1 $ is true; otherwise, $\chi_{ \{ j >1 \} } = 0$.
\end{theorem}
\begin{IEEEproof}
The proof is given in Appendix \ref{Pro:Exp_Waiting_Time_PQ}.
\end{IEEEproof}

\subsection{Calculation of  $\mathbb E \big[ {Z_j^T} \big]$}
During the transmission stage, the time spent for transmitting the processed data packet is directly related to both its size and the transmission rate. In particular, the instantaneous transmission rate is given by
\begin{align} \label{Eq:Rate_Random_Variable}
R_{D} = B{\log _2}\left( 1 + \frac{p_A h d^{ -\alpha}} {\sigma ^2} \right),
\end{align}
where $B$ is the bandwidth of the channel, $h$ is the channel power gain,  $d$ is the distance between the aggregator and destination node, and ${\sigma ^2}$ is the noise power.
Considering a Rayleigh fading, we note that the transmission rate $R_{D}$ is a random variable. Then, we derive the expected service time in the transmission subsystem  for the data packet  from each sensor  in the following theorem.

\begin{theorem} \label{Theorem:Exp_Transmission_Time_TQ}
The expected transmission time for successfully delivering a data packet originally generated by sensor $S_j$ is given as
\begin{align} \label{Eq:Transsmission_Time_Exp_Sj}
\mathbb E \big[ {Z_j^T} \big] =  {\frac{{{\xi _j}{\sigma ^2 d^\alpha }}}{p_A} \! \int_0^\infty \!\!\!\! \exp\! \left( {\frac{{{\xi _j}}}{t} + \frac{{ {{\rm{1}} - {\rm{exp}}\left( {\frac{{{\xi _j}}}{t}} \right)} }}{{{p_A} \sigma^{-2} {d^{ - \alpha }}}}} \right) \frac{dt}{t}},
\end{align}
where ${\xi _j} = {{{{\tilde C}_j}\ln 2}}/{B}$.
\end{theorem}
\begin{IEEEproof}
The proof is given in Appendix \ref{Pro:Exp_Transmission_Time_TQ}.
\end{IEEEproof}

\subsection{Calculation of  $\mathbb E \big[ {W_j^T} \big]$}
For the transmission subsystem, we can model it as a G/G/1 FCFS queueing system by considering that the inter-arrival time and service time follow different general distributions. The exact analytical results are usually unavailable for the G/G/1 queue, especially for the case where the distribution about the arrival or departure is unknown\cite{Queueing_Systems_1976}. For our concerned problem, the output of the processing queue is the input of the transmission queue. Although the average waiting time of update packets spent in the processing queue can be obtained according to Theorem 1, the high-order statistics for the inter-departure times of the processing queue are troublesome to derive due to the complexity. In this light, for the data transmission queue, it is inconvenient to get the distribution of the inter-arrival time of packets and further derive the expectation of the waiting time with classical approaches presented in \cite{Queueing_Systems_1976}.

In this subsection, we resort to implementing the principle of maximum entropy (PME) and getting an information theoretic approximation of the expected waiting time spent in the transmission queue. Concretely, we have the following theorem.

\begin{theorem} \label{Theorem:Exp_Waiting_Time_TQ}
In the transmission queue, the expectation of waiting time for packets from sensor $S_j$ can be mathematically approximated by
\begin{align}   \label{Eq:Waiiting_Time_Transmission_Q_Fin}
\mathbb E \big[ {W_j^T} \big] \thickapprox \frac{ {\mu _j} {{{\left( {\sum_{j = 1}^J {{\lambda _j}} \mathbb E \big[ {Z_j^T} \big] } \right)}^{\rm{2}}}}}{{\sum_{i = 1}^J {{\lambda _i}{\mu _i}} \left( {1 - \sum_{j = 1}^J {{\lambda _j}}\mathbb  E \big[  {Z_j^T} \big] } \right)}},
\end{align}
where ${\mu _1} =1$ and ${\mu _j}$ is given in (\ref{Eq:Waiiting_Time_Transmission_Q_Ratio}), $\forall j \in \{2,3,\cdots, J\}$, which represents the ratio of the expectation of waiting time for packets from sensor $S_j$ to that for packets from sensor $1$ in the transmission queue, i.e., ${\mu _j}=\frac{\mathbb E\big[ {W_j^T} \big]}{\mathbb E\big[ {W_1^T}\big]}$. In Eq. (\ref{Eq:Waiiting_Time_Transmission_Q_Ratio}), we have $H_{j,i}^P={\lambda _i}\left( {\mathbb E \big[ {W_i^P} \big] + \mathbb E \big[ {W_j^P} \big] } \right), \forall i \in \{1, 2, \cdots, j-1\}$, and $H_{j,j}^P = {\lambda _j}\mathbb E \big[ {W_j^P} \big]$.
\end{theorem}

\begin{figure*}
\begin{align}   \label{Eq:Waiiting_Time_Transmission_Q_Ratio}
\mu _j
&\approx \frac{{{P_{{A_P},B}}\mathbb E \big[ {{Z^T}} \big] + max\left\{{\sum_{i = 1}^j {H_{j,i}^P}\mathbb E \big[ {Z_i^T} \big] - (\mathbb E \big[ {W_j^P} \big]+\mathbb E \big[ {Z_j^P} \big])}, 0 \right\}}}{{{P_{{A_P},B}}\mathbb E \big[ {{Z^T}} \big] +max\left\{ {H_{1,1}^P}\mathbb E \big[ {Z_1^T} \big]-(\mathbb E \big[ {W_1^P} \big]+\mathbb E \big[ {Z_1^P} \big]),0 \right\}}}
\end{align}
\hrulefill
\end{figure*}

\begin{IEEEproof}
The proof is given in Appendix \ref{Pro:Exp_Waiting_Time_TQ}.
\end{IEEEproof}

\begin{remark}
We note Eq. (\ref{Eq:Waiiting_Time_Transmission_Q_Ratio}) implicitly means that for a typical packet, its expected waiting time in the transmission queue is approximately proportional to the difference between two parts: 1. The overall transmission time of packets processed by the processor when it waited for its service in the processing queue; 2. Its experienced system time (i.e., the sum of waiting time and service time) in the data processing system. More details can be found in Appendix \ref{Pro:Exp_Waiting_Time_TQ}.
\end{remark}

Based on the analyses made in the above three subsections, we can obtain the expression of the average PAoI for packets from each sensor $S_j$ by combining Eq. (\ref{Eq:Average_PAOI_Detail}), (\ref{Eq:Remaining Processing_Time_Fin}), (\ref{Eq:Transsmission_Time_Exp_Sj}), and (\ref{Eq:Waiiting_Time_Transmission_Q_Fin}).

\subsection{Validation}
\begin{table}[!t]
\renewcommand{\arraystretch}{1.3}
\caption{Default Simulation Parameters}
\label{Tab:Parameters}
\centering
\begin{tabular}{ l  l  } \toprule
\textbf{Parameter and description} & \textbf{{value}} \\
\midrule
Number of sensors, $J$                                                                                          & 3                                                 \\
Size of original packets, $\{C_1, C_2, C_3\}$                                                             & $\{10, 7, 5\}$ Mbits                           \\
Size of processed packets, $\{\tilde C_1, \tilde C_2, \tilde C_3\}$                                       & $\{2, 2, 2\}$ Mbits                            \\
Transmit power of the aggregator, $p_A$                                                                         & 100 mW                                            \\
Distance between A and D, $d$                                                   & 300 m                                             \\
The transmission bandwidth, $B$                                                                                     & 100 KHz                                             \\
The path loss exponent, $\alpha$                                                                                 & 3                                                 \\
The AWGN power density,                                                                                             & -174 dBm/Hz                                       \\
\toprule
\end{tabular}
\end{table}

\begin{figure*}[!t]
\centering
\subfigure[]{\centering \includegraphics[width=2.35in]{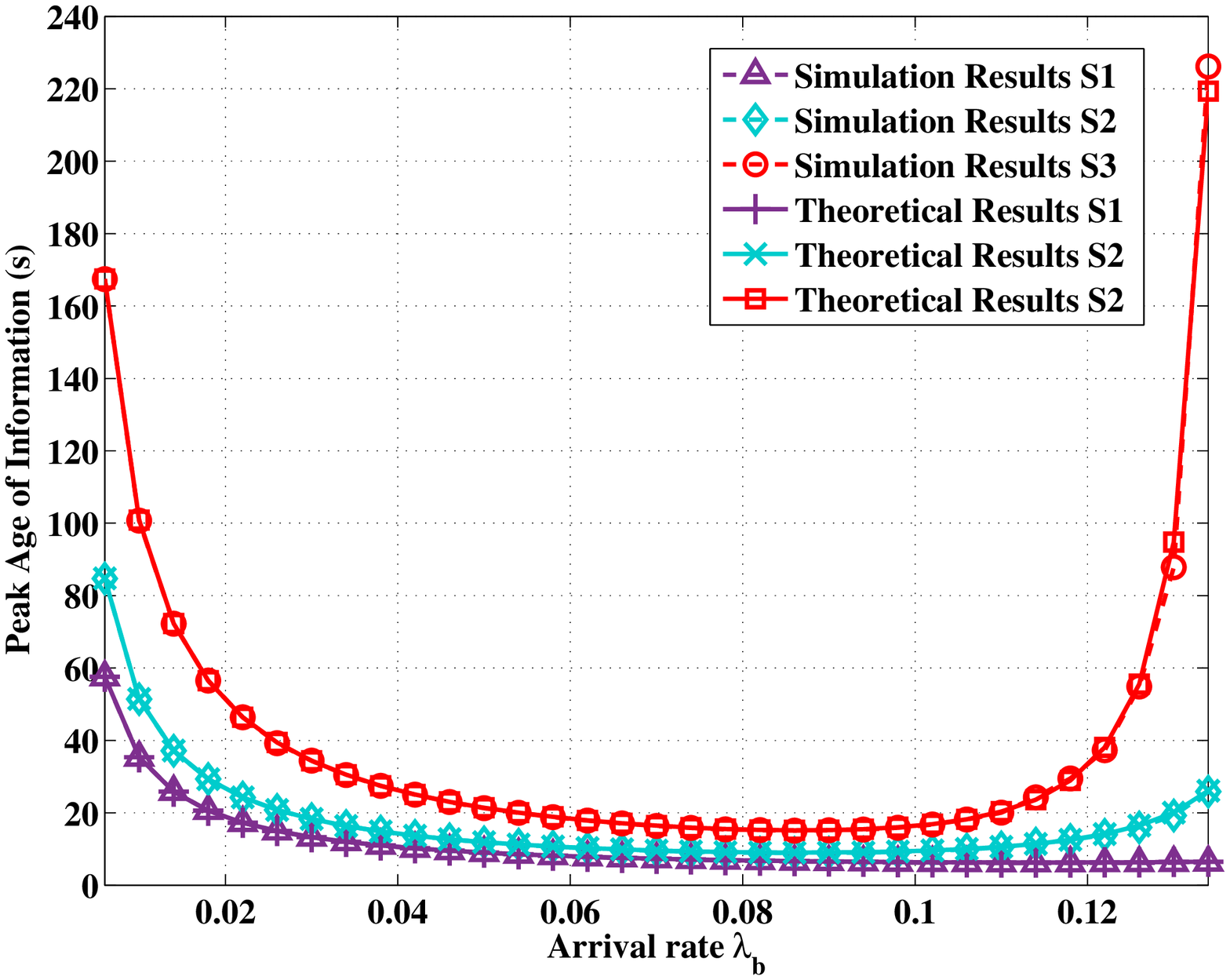}}%
\subfigure[]{\centering \includegraphics[width=2.35in]{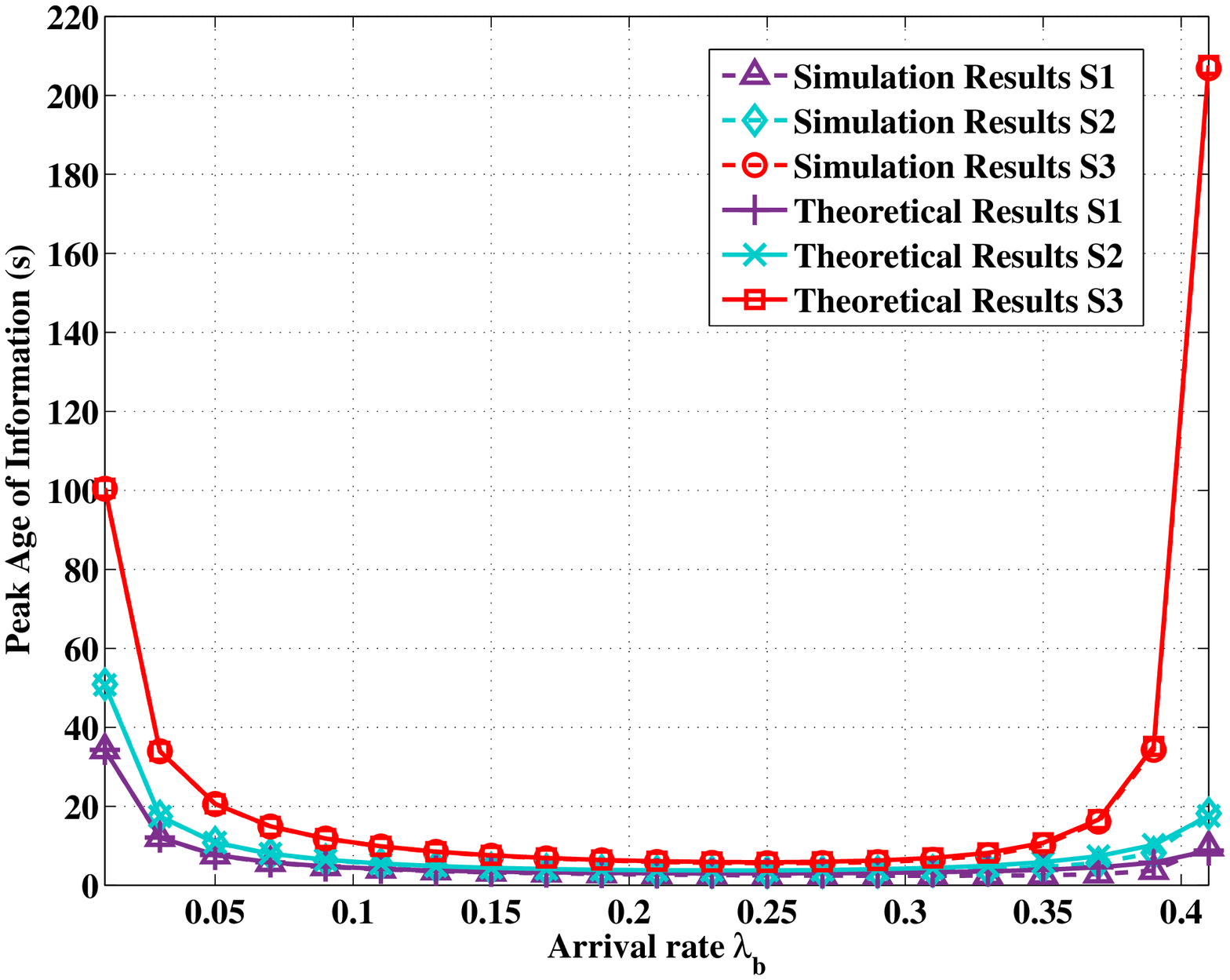}}%
\subfigure[]{\centering \includegraphics[width=2.35in]{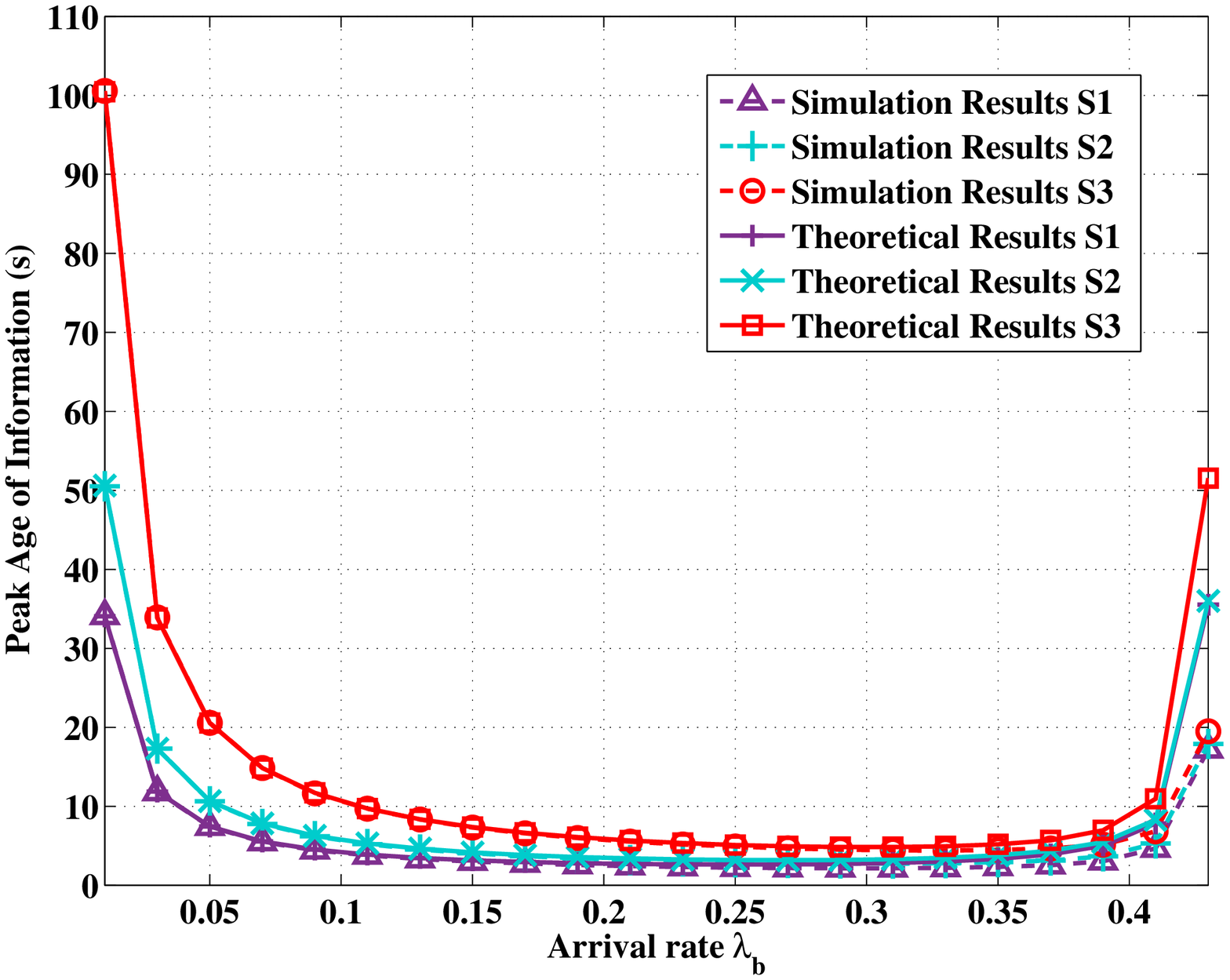}}%
\centering
\centering \caption{Average PAoI for packets from different sensors, where the equivalent processing rate, $\frac {r}{\tau _j}$, $\forall j \in \{1, 2, 3\}$ is: (a) 5 Mbits/s, i.e., $\mathbb E\big[ {Z^P} \big] = 1.200 $ and $\mathbb E\big[ {{Z^T}} \big] = 0.384$; (b) 15 Mbits/s, i.e., $\mathbb E\big[ {Z^P} \big] = 0.400$ and $\mathbb E\big[ {{Z^T}} \big] = 0.384$; (c) 25 Mbits/s, i.e., $\mathbb E\big[ {Z^P} \big] = 0.240$ and $\mathbb E\big[ {{Z^T}} \big] = 0.384$;.} \label{Fig:Theoretic_vs_Simulation}
\end{figure*}

We now verify the accuracy of our analysis via simulations. Specifically, we set the arrival rate of update packets from sensor $S_j$ as $\lambda _j = (J-j)\lambda _b$, where $\lambda _b$ can be regarded as the \textit{basic} arrival rate with the units packets/second. This setting means that the packets with a higher processing priority would also have a larger arrival rate. Other parameters are set according to Table~\ref{Tab:Parameters}. In this subsection, all the simulation results are obtained by averaging over $10^6$ realizations, i.e., by averaging the PAoI for $10^6$ update packets. We note that by introducing the \textit{basic} arrival rate $\lambda_b$, the comparison of simulation and theoretical results for different sensors can be clearly presented in one figure, as shown in Fig. \ref{Fig:Theoretic_vs_Simulation}.

Fig. \ref{Fig:Theoretic_vs_Simulation}(a) compares the simulation and theoretical results on the average PAoI for different sensors, where the equivalent processing rate, $\frac {r}{\tau _j}$, $\forall j \in \{1, 2, 3\}$, is 5 Mbits/s, and the data processing time dominates transmission time, i.e., $\mathbb E \big[ {Z^P} \big] =  1.200$ s and $\mathbb E\big[ {{Z^T}} \big] = 0.384$ s. The results show a close match for all sensor, which validate the our mathematical analysis. It can be observed from Fig. \ref{Fig:Theoretic_vs_Simulation}(a) that even the traffic load is high, e.g., $\lambda _b = 0.13$, the average PAoI for packets from sensor $S_1$ is still kept low, i.e., about 6.4 s, while those for sensor $S_2$ and $S_3$ are about 20.1 s and 94.8 s, respectively. This is due to the fact that the priority based processing subsystem will try its best to first provide required service to update packets with the highest priority even in the case where the resource is not enough to provide good service to all sensors. In other words, the average PAoI for the packets with the lowest priority will first suffer significant performance degradation as the traffic load becomes heavy, while the packets with the highest priority would be delivered timely. Hence, to make the whole system working in a stable state we need to properly control the packet arrival rates for all sensors, which will be presented and studied in detail in Section 4.

To see what happens when the data processing time is comparable to or even dominated by the data transmission time, we respectively set the equivalent processing rate, $\frac {r}{\tau _j}$, $\forall j \in \{1, 2, 3\}$ to 15 Mbits/s (i.e., $\mathbb E \big[{Z^P} \big] = 0.400$ and $\mathbb E\big[ {{Z^T}} \big] = 0.384$) and 25 Mbits/s (i.e., $\mathbb E \big[{Z^P} \big] = 0.240$ and $\mathbb E\big[ {{Z^T}} \big] = 0.384$) in Fig. \ref{Fig:Theoretic_vs_Simulation}(b) and Fig. \ref{Fig:Theoretic_vs_Simulation}(c). In these two figures, the simulation results match well with the theoretical results in the low traffic load region, while, in the heavy load region, the theoretic results is slightly higher than the simulation results. This is mainly due to the fact when the transmission subsystem gradually acts as the bottleneck, the waiting time in the transmission queue brings more obvious effects to the average PAoI. However, according to the proof of Theorem \ref{Theorem:Exp_Waiting_Time_TQ}, for the transmission queue, only the first moment of the number of waiting packets is incorporated when adopting the principle of maximum entropy, which makes the obtained average waiting time undervalued, i.e., higher than that in practice. This could be more obviously observed in Fig. \ref{Fig:Theoretic_vs_Simulation}(c) where the data transmission time dominates the data processing time. Particularly, as the load increase, the waiting time gradually beats the inter-arrival time and plays a pivotal role in determining the information freshness, and the difference between the simulation results and theoretical results becomes larger. Nevertheless, the variation trend of the average PAoI could be well captured by our theoretic analysis. Besides, by combining Fig. \ref{Fig:Theoretic_vs_Simulation}(a)-(c), it can be seen that, given the arrival rate $\lambda _b$, the advantage brought by data preprocessing is more significant when the processing rate is higher. This is because, for the same $\lambda _b$, the system time could be decreased if the aggregator is equipped with a faster data processor, since the time spent in the data processing queue is reduced.

In conclusion, it's feasible to obtain the average PAoI and further derive the optimal data arrival rates for distinct sensors with our theoretical analysis presented in this section, when the processing time is dominating or comparable with the transmission time. On the other hand, for the case that the processing time is extremely short 
and there is no queueing for the processing subsystem, our formulated tandem queue (Fig. \ref{Fig:Fig1}) degrades into a M/G/1 queue where the packet arrival from each individual sensor is delayed for a fix time for data processing. Then, it can be regarded as one special case that studied in available work \cite{Multi_Source_ISIT,Multi_Source_ARXIV}.

\section{Problem Formulation and Algorithm Design} \label{Sec:PAoI_Optimization}
The analysis derived in Section~\ref{Sec:PAoI_Analysis} allows us to perform further control on the update frequency so as to optimize the information freshness. Specifically, using the analytical expressions, we can formulate the problem into the following min-max programming \cite{Multi_Source_ISIT,Max_Min_PAoI}
\begin{eqnarray}   \label{Eq:PAoI_Min_Max_p}
\textbf{P}: &\mathop { \min}\limits_{ {\mathbf {\Lambda}} }   & \max \left\{ {{A _j}\left| \, {j \in \left\{ {1,2,\cdots,J} \right\}} \right.} \right\} \\ \label{Eq:P_Min_Max}
&\textrm{s.t.} & \sum\limits_{j = 1}^J {\lambda _j} {\mathbb E \big[ {Z_j^P} \big] } < 1, \label{Eq:P_PQ_Stable_Constraint} \\
&& \sum\limits_{j = 1}^J {\lambda _j} {\mathbb E \big[ {Z_j^T} \big] }  < 1, \label{Eq:P_TQ_Stable_Constraint} \\
&& \lambda _j >0, ~~ \forall j \in \mathcal J  \label{Eq:Nonnegative_Constraint}
\end{eqnarray}
where $ {\mathbf {\Lambda}} = (\lambda_1, \lambda_2, \cdots, \lambda_J)$ represents the vector of update arrival rates and ${A _j}$ denotes the average PAoI for updates from sensor $S_j$. The constraints (\ref{Eq:P_PQ_Stable_Constraint}) and (\ref{Eq:P_TQ_Stable_Constraint}) guarantee the stability of the processing and transmission subsystems, respectively. Additionally, the non-negativity of each individual variable is guaranteed with constraint (\ref{Eq:Nonnegative_Constraint}). Here, we consider the case that the above two constraints (\ref{Eq:P_PQ_Stable_Constraint}) and (\ref{Eq:P_TQ_Stable_Constraint}) are linearly independent, i.e., $\left(\mathbb E\big[ {Z_1^P}\big],\mathbb E\big[ {Z_2^P} \big],\cdots,\mathbb E\big[ {Z_J^P} \big]\right) \neq \varsigma \left(\mathbb E\big[ {Z_1^T} \big],\mathbb E\big[{Z_2^T} \big],\cdots,\mathbb E\big[ {Z_J^T} \big]\right)$ where $\varsigma \in \mathbb R$ is a constant. Otherwise, only the tighter one needs to be addressed and the problem degenerates into that with $J+1$ constraints.

While problem \textbf{P} looks simple as incorporating only linear constraints, solving it is actually non-trivial. Because the common processing and transmission resources are shared by different sensors, their achieved average PAoIs are closely correlated and have no closed-form expression. Hence, the convexity of this problem can not be readily established and traditional efficient algorithms resorting to the derivatives of the object function is prohibitive to be applied.
In this regard, we adopt the derivative-free optimization methods\cite{Introduction_DO_Book,DO_Review_No_Constraint} to address our formulated problem \textbf{P}.

Motivated by the recent work \cite{GSS_Linearly_Constraint_Opt}, we develop a Generating set search based Average PAoI minimization (GAP) algorithm to solve problem \textbf{P}. In general, generating set search, introduced in \cite{GSS_Org}, is one of frameworks for globally convergent directional direct-search methods, with which some feasible points (trial points) are selected and evaluated in each iteration, and the solution is finally obtained when the stop criterion is satisfied. For each optimization programming with constraints, to guarantee the global convergence\footnote{For one algorithm, the global convergence means that a stationary point can be finally achieved from an arbitrarily chosen starting point.} of the developed generating set search based algorithm, two key issues have to be well addressed in each iteration: 1. how to select a suitable set of searching directions; 2. how to choose an appropriate step-size for each individual searching direction. Next, we would provide clues about how to address these two issues for our concerned problem in Section V-A, present the details of the devised algorithm as well as its convergence analysis in Section V-B, and finally give the numerical results and discussions in Section V-C.

\subsection{Preliminary on GAP Algorithm Design}

Here, based on the problem \textbf{P}, we introduce some preliminary knowledge about the principle of selecting available searching directions and corresponding step-sizes, which is the fundamental basis of the further algorithm design in the following subsection. For the sake of presentation, we first rewrite problem \textbf{P} as the following equivalent problem:
\begin{eqnarray}   \label{Eq:PAoI_Min_Max_p1}
\textbf{P1}: &\mathop {\min}\limits_{ {\mathbf {\Lambda}} }    & \widetilde{A}\left({\mathbf {\Lambda}}\right)\\ \label{Eq:P1_Min_Max}
&\textrm{s.t.} & \mathbf {Z}{\mathbf {\Lambda}^T} < \mathbf b ^T \label{Eq:P1_Matrix_Constraint}
\end{eqnarray}
with $\widetilde{A}\left({\mathbf {\Lambda}}\right) = \max \left\{ {{A _j}\left| {j \in \left\{ {1,2,\cdots,J} \right\}} \right.} \right\} $ and
\begin{align} \label{Eq:Matrx_A}
\mathbf Z = {\left[ {\begin{array}{*{20}{c}}
{{\mathbf{z}}_1^\mathbf T}&{{\mathbf{z}}_2^\mathbf T}&{{\mathbf{z}}_3^\mathbf T}& \cdots &{{\mathbf{z}}_{J{\rm{ + }}2}^\mathbf T}
\end{array}} \right]^\mathbf T}= \left[ {\begin{array}{*{20}{c}}
\mathbf Z_1 \\
-\mathbf I
\end{array}} \right].
\end{align}
Wherein, $\mathbf Z$ denotes a $J \times (J+2)$ parameter matrix associated to the $J+2$ inequality constraints presented in Eq. (\ref{Eq:P_PQ_Stable_Constraint})-(\ref{Eq:Nonnegative_Constraint}), and $(\cdot)^{\mathbf T}$ denotes the transpose operation. Particularly, the matrix $\mathbf Z_1$ and vector $\mathbf b $ can be expressed as
\begin{align} \label{Eq:Matrx_A1}
\mathbf Z_1 = \left[ {\begin{array}{*{20}{c}}
{\mathbb E\big[ {Z_1^P} \big]}&{\mathbb E\big[  {Z_2^P} \big]}& \cdots &{\mathbb E\big[  {Z_J^P} \big]}\\
{\mathbb E\big[ {Z_1^T} \big]}&{\mathbb E\big[  {Z_2^T} \big]}& \cdots &{\mathbb E\big[  {Z_J^T} \big]}
\end{array}} \right]
\end{align}
and
\begin{align} \label{Eq:Vector_b}
\mathbf b  &= \big[ {\begin{array}{*{20}{c}}\!b_1\!&\!b_2\!&\!b_3\!&\!\cdots\!&b_{J+2} \!\end{array}} \big] = \big[ {\begin{array}{*{20}{c}}\!1\!&\!1\!&\!0\!&\!\cdots\!&0\!\end{array}} \big]
\end{align}
respectively, and $\mathbf I$ is a $J \times J$ unit matrix.

Hereafter, we focus on solving \textbf{P1} and introduce some basic notations and definitions as follows. Let $\Omega  = \left\{ {{\bf{\Lambda }}\left| {{\bf{Z\Lambda ^T}} < {\bf{b}^T}} \right.} \right\}$ and ${\Upsilon _l} = \left\{ {{\bf{\Lambda }}\left| {\mathbf z_l{\bf{\Lambda }^T} = {b_l}} \right.} \right\}$ denote the feasible region and set where the $l$-th constraint is (virtually) binding, respectively. Given a feasible point $\mathbf{\Lambda } \in \Omega$, the set of indexes of the $\varepsilon$-binding constraints is defined as \cite{GSS_Linearly_Constraint_Opt}
\begin{align}  \label{Eq:Var_Bind_Con_Ind}
\mathcal I\left( {{\bf{\Lambda }},\varepsilon } \right) = \left\{ {l\left| {D\left( {{\bf{\Lambda }},{\Upsilon _l}} \right) \le \varepsilon } \right.} \right\}
\end{align}
where $D( {{\bf{\Lambda }},{\Upsilon _l}} )$ represents the distance from $\mathbf{\Lambda } \in \Omega$ to the boundary face of $\Omega$ according to the $l$-th constraint.

Accordingly, each vector with the index belonging to $\mathcal I( {{\bf{\Lambda }},\varepsilon } )$ is the outward-pointing normal to one boundary face of $\Omega$ which is within distance $\varepsilon$ from point ${\bf{\Lambda }}$. Meanwhile, for one feasible point $\mathbf{\Lambda } \in \Omega$, we can define the $\varepsilon$-normal cone $\mathcal N({{\bf{\Lambda }},\varepsilon })$ to be the cone generated by the set of vectors $\left\{ {{\mathbf z_l}\left| {l \in \mathcal I( {{\bf{\Lambda }},\varepsilon } )} \right.} \right\} \cup \left\{ \mathbf 0 \right\}$, i.e.,
\begin{align}  \label{Eq:Var_Cone}
&\mathcal N\left( {{\bf{\Lambda }},\varepsilon } \right) \\ \nonumber
&=\begin{cases}
\left\{ {\sum\limits_{l \in \mathcal I\left( {{\bf{\Lambda }},\varepsilon } \right)} {{\varpi _l}{\mathbf z_l}} \left| {{\varpi _l} \ge 0,l \in \mathcal I\left( {{\bf{\Lambda }},\varepsilon } \right)} \right.} \right\}, &\mathcal I\left( {{\bf{\Lambda }},\varepsilon } \right) \ne \emptyset \\
\left\{ \mathbf 0 \right\}, &\mathcal I\left( {{\bf{\Lambda }},\varepsilon } \right) = \emptyset
\end{cases}
\end{align}
where $\left\{ \mathbf 0 \right\}$ and $\emptyset$ represent a $1 \times J$ zero vector and the empty set, respectively. Furthermore, we coin the term $\varepsilon$-tangent cone $\mathcal T({{\bf{\Lambda }},\varepsilon } )$ for the polar of the cone $\mathcal N( {{\bf{\Lambda }},\varepsilon} )$, i.e,
\begin{align}  \label{Eq:Var_Polar_Cone}
\mathcal T\left( {{\bf{\Lambda }},\varepsilon } \right) = \left\{ {\mathbf x\left| {\mathbf x{\mathbf y^T} \le 0,\forall \mathbf y \in \mathcal N\left( {{\bf{\Lambda }},\varepsilon } \right)} \right.} \right\}
\end{align}
where each element $\mathbf x$ in set $\mathcal T( {{\bf{\Lambda }},\varepsilon } )$ is a $1 \times J$ row vector. It should be noted that if $\mathcal N( {{\bf{\Lambda }},\varepsilon } ) = \mathbb{R}^J$ then $\mathcal T( {{\bf{\Lambda }},\varepsilon } ) = \{ \mathbf 0 \}$. Meanwhile, if $\mathcal N( {{\bf{\Lambda }},\varepsilon } ) = \{ \mathbf 0 \}$ then $\mathcal T( {{\bf{\Lambda }},\varepsilon }) = \mathbb{R}^J$.

\begin{figure}[!t]
\centering \includegraphics[width=3.0in]{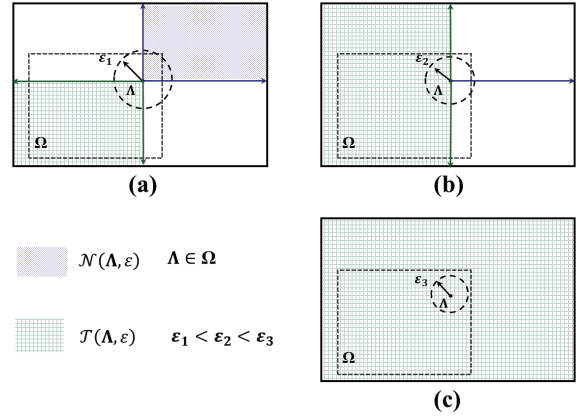}
\centering \caption{An example illustration of sets $\mathcal N( {{\bf{\Lambda }},\varepsilon } ) $ and $\mathcal T( {{\bf{\Lambda }},\varepsilon } ) $ in $\mathbb R^2$ space, where three different cases with distinct values $\varepsilon _1$, $\varepsilon _2$ and $\varepsilon _3$ ($\varepsilon _1 < \varepsilon _2 < \varepsilon _3$) are presented. We note that when $\varepsilon$ is small enough (e.g., $\varepsilon _3$) then $\mathcal N( {{\bf{\Lambda }},\varepsilon } ) = \{ \mathbf 0 \}$ and $\mathcal T( {{\bf{\Lambda }},\varepsilon } ) = \mathbb{R}^2$. } \label{Fig:Normal_Polar_Cones_Illustration}
\end{figure}

In fact, if $\mathcal T( {{\bf{\Lambda }},\varepsilon } ) \ne \{ \mathbf 0 \}$, then from the point ${\bf{\Lambda }}$ and along all the directions specified by $\mathcal T( {{\bf{\Lambda }},\varepsilon } )$, other feasible points (i.e., staying in the feasible region) can be achieved, and the distance between $\mathbf \Lambda$ and each feasible point is not greater than $\varepsilon$.\footnote{This proposition is always true for the case that the feasible region is specified by linear constraints \cite{GSS_Linearly_Constraint_Opt}.} To ease the understanding,  we present an example illustration of sets $\mathcal N( {{\bf{\Lambda }},\varepsilon } ) $ and $\mathcal T( {{\bf{\Lambda }},\varepsilon } ) $ in $\mathbb R^2$ space in Fig. \ref{Fig:Normal_Polar_Cones_Illustration}, where the whole feasible region is a rectangle with dashed edges labeled by $\Omega$. As shown in Fig. \ref{Fig:Normal_Polar_Cones_Illustration}, as $\varepsilon$ varies from $\varepsilon _1$ to $\varepsilon _3$, the size of set $\mathcal N( {{\bf{\Lambda }},\varepsilon } ) $ gradually becomes small while that of $\mathcal T( {{\bf{\Lambda }},\varepsilon } ) $ is larger. According to this trend, given ${\bf{\Lambda }}$ as the starting point and $\varepsilon$ as the upper bound of the step-size for searching, more directions endow us feasible points lie in $\Omega$. We note that if $\varepsilon$ is small enough (e.g., $\varepsilon _3$) then $\mathcal N( {{\bf{\Lambda }},\varepsilon } ) = \{ \mathbf 0 \}$ and $\mathcal T( {{\bf{\Lambda }},\varepsilon } ) = \mathbb{R}^2$, i.e., there is no infeasible points lie within the distance less than $\varepsilon _3$ from the starting feasible point ${\bf{\Lambda }}$.

As discussed above, starting from any feasible point ${\bf{\Lambda }}$, the suitable searching directions and step-sizes could be determined based on the $\varepsilon$-tangent cone $\mathcal T({{\bf{\Lambda }},\varepsilon} )$ and its parameter $\varepsilon$, respectively. The developed algorithm GAP and its convergence analysis will be presented in detail in the following subsection.

\subsection{GAP Algorithm for Solving Problem \textbf{P1}}

Based on the preliminary presented in the previous subsection, here we develop GAP algorithm to solve problem \textbf{P1} (equivalent to the original problem \textbf{P}) as shown in Algorithm \ref{Alg:GAP_algorithm}, where ${\mathbf{\Lambda } (t)}$ denotes the feasible solution ${\bf{\Lambda }}$ chosen in the $t$-th iteration. In this algorithm, the loop is repeated until the searching step-size is small enough, i.e., $0< {\Phi (t)} < {\Phi _{\min}} $ with ${\Phi _{\min}} \rightarrow 0_{+}$ (e.g., ${\Phi _{\min}} = 10^{-5}$).

\begin{algorithm}
\caption{Generating set search based Average PAoI minimization (GAP) algorithm.}
\begin{algorithmic}[1] \label{Alg:GAP_algorithm}
\STATE \textbf{Initialization:}
\STATE Set $t=0$ and ${\Phi (t)}$ with Eq. (\ref{Eq:Initial_Potential_Step_Size}) as the initial value of the potential step-size for searching. Randomly generate a point ${\mathbf{\Lambda } (t)} \in \Omega$ as the initial guess.
\STATE \textbf{Go into a loop:}
\STATE Set $\mathbf{\Lambda } = {\mathbf{\Lambda } (t)}$ and $\varepsilon = \min \{\varepsilon_{max}, {\Phi (t)}\}$ . \label{Alg1:Iteration_Beg}
\STATE \textbf{Searching directions and step-sizes generation:}
\STATE Derive the set of candidate searching directions $\mathcal S_{\mathcal T( {{\mathbf{\Lambda }},\varepsilon } )}$ with Eq. (\ref{Eq:Polar_Cone_Generators}). Adopt $\Phi _{\mathbf s_l}$ (with Eq. (\ref{Eq:Searching_Step_Size})) as the searching step-size for each direction $\mathbf s_l$.
\STATE \textbf{Evaluation for trial points:}
\STATE Set ${\mathbf{\Lambda } (t+1)} = \bf{\Lambda }$ and put $\bf{\Lambda }$ into the candidate set $\mathcal O _t$.
\FOR {Each direction $\forall \mathbf s_l \in \mathcal S_{\mathcal T( {{\bf{\Lambda }},\varepsilon } )}$}
\IF{$\widetilde{A}({\mathbf \Lambda {+{\Phi _{\mathbf s_l} \mathbf s_l}}}) < \widetilde A ( {\bf{\Lambda }} ) - {{10}^{ - 4}}{{( {{\Phi _{{{\bf{s}}_l}}}} )}^2}$}
\STATE Put ${\check{\mathbf{\Lambda }} _{\mathbf s_l}}= {\mathbf \Lambda+{\Phi _{\mathbf s_l} \mathbf s_l}}$ into the candidate set $\mathcal O _t$.
\ENDIF
\ENDFOR
\STATE Set ${\mathbf{\Lambda } (t+1)}= \mathop {\arg }\limits_{{\check{\mathbf{\Lambda }} _{\mathbf s_l}}} \min \{ {\widetilde A ( {{\check{\mathbf{\Lambda }} _{\mathbf s_l}}})\left| {{\forall{\check{\mathbf{\Lambda }} _{\mathbf s_l}}} \in {\mathcal O_t}} \right.} \}$. \label{Alg1:Choosing_Opt_Point}
\STATE \textbf{Potential searching step-size update:}
\IF{${\mathbf{\Lambda } } \neq {\mathbf{\Lambda } \left(t+1\right)} $}
\STATE Set ${\Phi (t+1)} = {\Phi (t)}$.  \label{Alg1:Update_Opt_Point}
\ELSE
\STATE Set ${\Phi (t+1)} = \frac{1}{2}{\Phi (t)}$. \label{Alg1:No_Update_Opt_Point}
\ENDIF
\STATE \textbf{Termination checking:}
\IF{${\Phi (t)} < {\Phi _{\min}}$}
\STATE Go to \ref{Alg1:Iteration_End}.
\ELSE
\STATE Set $t=t+1$ and go to \ref{Alg1:Iteration_Beg}.
\ENDIF
\STATE \textbf{Output:} $\mathbf{\Lambda }^* = \mathbf{\Lambda } (t+1)$.
\label{Alg1:Iteration_End}
\end{algorithmic}
\end{algorithm}

At the beginning of GAP, the starting point (initial guess) ${\mathbf{\Lambda } (0)}$ is randomly selected from the feasible region $\Omega$. Meanwhile, the potential step-size for searching is initialized as ${\Phi (0)}>{\Phi _{\min}} >0$. For instance, we could set it as
\begin{align}  \label{Eq:Initial_Potential_Step_Size}
{\Phi \left(0\right)}\! = \!{\max\! \left\{\!{\left\{ {{\frac{1}{\mathbb E \big[ {Z_j^P} \big]}}\left| {j \in \mathcal J} \right.} \!\right\} \!\cup \!\left\{ {{\frac{1}{\mathbb E \big[ {Z_j^T} \big]}}\left| {j \in \mathcal J}\! \right.} \right\}}\! \right\}}\!.
\end{align}
After that, the algorithm goes into a loop. At each iteration $t$, with the current feasible point ${\mathbf{\Lambda } (t)}$ as the origin, we will first determine the candidate searching directions and corresponding step-sizes. Then, we calculate the function values for generated trail points and get the best solution obtained after this iteration. In the sequel, we introduce these two parts in detail.

As presented in the previously subsection, we note that staring from a feasible point ${\bf{\Lambda }}$ and searching along any direction (vector) in the $\varepsilon$-tangent cone $\mathcal T( {{\bf{\Lambda }},\varepsilon } ) $, we can always get a feasible point with the step-size less than $\varepsilon$. Hence, if $\mathcal T( {{\bf{\Lambda }},\varepsilon } ) $ is $\{ \mathbf 0\}$ then only $\{ \mathbf 0\}$ is the feasible direction, i.e., no other feasible points can be find starting from ${\bf{\Lambda }}$. In addition, if $\mathcal T( {{\bf{\Lambda }},\varepsilon } ) = \mathbb R^J$ then, as for the traditional programming without constraints, the set of candidate searching directions could be a positive basis in $\mathbb R^J$, e.g., $\mathcal {P} = \{\mathbf e_1, \mathbf e_2, \cdots, \mathbf e_J, -\mathbf e_1, -\mathbf e_2, \cdots, -\mathbf e_J\}$, where $\mathbf e_j$ denotes a $1 \times J$ row vector with the $j$-th element being 1 and other elements being 0.\footnote{We note that a positive spanning set in $\mathbb R^J$ is a set of vectors whose cone is $\mathbb R^J$. Moreover, if every vector in a positive spanning set is positively independent of others, this positive spanning set is a positive basis for $\mathbb R^J$. In fact, for $\mathbb R^J$ the positive basis is not unique and $\mathcal {P}$ is a maximal positive basis. Please see \cite{Introduction_DO_Book} for details.} In cases $\mathcal T( {{\bf{\Lambda }},\varepsilon })$ is neither $\mathbb R^J$ nor $\{ \mathbf 0 \}$, the set of candidate searching directions $\mathcal S_{\mathcal T( {{\bf{\Lambda } },\varepsilon } )}$ could be constructed with a set of generators of $\mathcal T( {{\bf{\Lambda }},\varepsilon } )$.
\begin{proposition}\label{Pro:Generator}
Denote $\mathbf F$ as a matrix whose rows consist of the linearly independent generators of $\mathcal N( {{\bf{\Lambda }},\varepsilon } )$, and $\mathbf D$ as a matrix whose rows constitute a positive basis for the null-space of $\mathbf F^T$. Then, we have the matrix
\begin{align}  \label{Eq:Polar_Cone_Generators_Matrix}
\mathbf S = \left[ {\begin{array}{*{20}{c}}
{\bf{D}}\\
{ - \mathbf F{{\left( {{\mathbf F^T}\mathbf F} \right)}^{ - 1}}}
\end{array}} \right]
= \left[ {\begin{array}{*{20}{c}}
{{\bf{I}} - {{\bf{F}}^T}{{\left( {{\bf{F}}{{\bf{F}}^T}} \right)}^{ - 1}}{\bf{F}}}\\
{{{\bf{F}}^T}{{\left( {{\bf{F}}{{\bf{F}}^T}} \right)}^{ - 1}}{\bf{F}} - {\bf{I}}}\\
{ - {{\left( {{\bf{F}}{{\bf{F}}^T}} \right)}^{ - 1}}{\bf{F}}}
\end{array}} \right]
\end{align}
whose rows are the generators of the polar cone $\mathcal T( {{\bf{\Lambda }},\varepsilon } )$. Wherein, $\mathbf I$ denotes a $J \times J$ unit matrix.
\end{proposition}
\begin{IEEEproof}
Readers are referred to the proof of Proposition 8.2 in \cite{Proposition_Proof_Ref} for the similar processes by just taking $\varepsilon$ as the maximum allowed distance from the feasible point to constraints in each iteration. The details are omitted here due to space limitations.
\end{IEEEproof}

To construct the matrix $\mathbf F$, one promising and feasible option is to set its rows as the elements in $\{ {{\mathbf z_l}\left| {l \in \mathcal I( {{\bf{\Lambda }},\varepsilon } )} \right.} \}$. Then, according to Proposition \ref{Pro:Generator}, $\mathcal S_{\mathcal T( {{\bf{\Lambda } },\varepsilon } )}$ can be set as follows
\begin{align}  \label{Eq:Polar_Cone_Generators}
&\mathcal S_{\mathcal T\left( {{\bf{\Lambda } },\varepsilon } \right)} \!=\!\begin{cases}\!
\mathcal {P}, &\mathcal T\left( {{\bf{\Lambda } }\!,\!\varepsilon } \right) = \mathbb R^J \\
\!\left\{ \mathbf 0 \right\}, & \mathcal T\left( {{\bf{\Lambda } }\!,\!\varepsilon } \right)  = \emptyset \\
\!\left\{ \mathbf s_1, \!\mathbf s_2, \!\cdots, \!\mathbf s_{2J+\left| {\mathcal I( {{\bf{\Lambda }},\varepsilon })} \right|} \right\}, & \text{otherwise}
\end{cases}
\end{align}
where $\mathbf s_l$, $\forall l = \{1, 2, \cdots, {2J+\left| {\mathcal I( {{\bf{\Lambda }},\varepsilon } )} \right|}\}$, denotes the $l$-th row of the matrix $\mathbf S $ in Eq. (\ref{Eq:Polar_Cone_Generators_Matrix}), i.e.,
\begin{align}  \label{Eq:Polar_Cone_Generators_Matrix_Vector_Form}
\mathbf S = {\left[ {{\mathbf {s}}_1^\mathbf T,{\mathbf {s}}_2^\mathbf T, \cdots ,{\mathbf {s}}_{{2J+\left| {\mathcal I\left( {{\bf{\Lambda }},\varepsilon } \right)} \right|}}^\mathbf T} \right]^\mathbf T},
\end{align}
and $| {\cdot} |$ represents the cardinality of a set. Accordingly, along each direction $\mathbf s_l$ the adopted searching step-size is set as
\begin{align}  \label{Eq:Searching_Step_Size}
{\Phi _{\mathbf s_l} } = \frac{ \varepsilon} {\left\| {{{\bf{s}}_l}} \right\|}, \forall \mathbf s_l \in \mathcal S_{\mathcal T\left( {{\bf{\Lambda } },\varepsilon } \right)}.
\end{align}
where $\|  {\cdot}  \|$ denotes the Euclidean norm of a vector.

\begin{figure*}[!t]
\centering
\subfigure[]{\centering \includegraphics[width=2.35in]{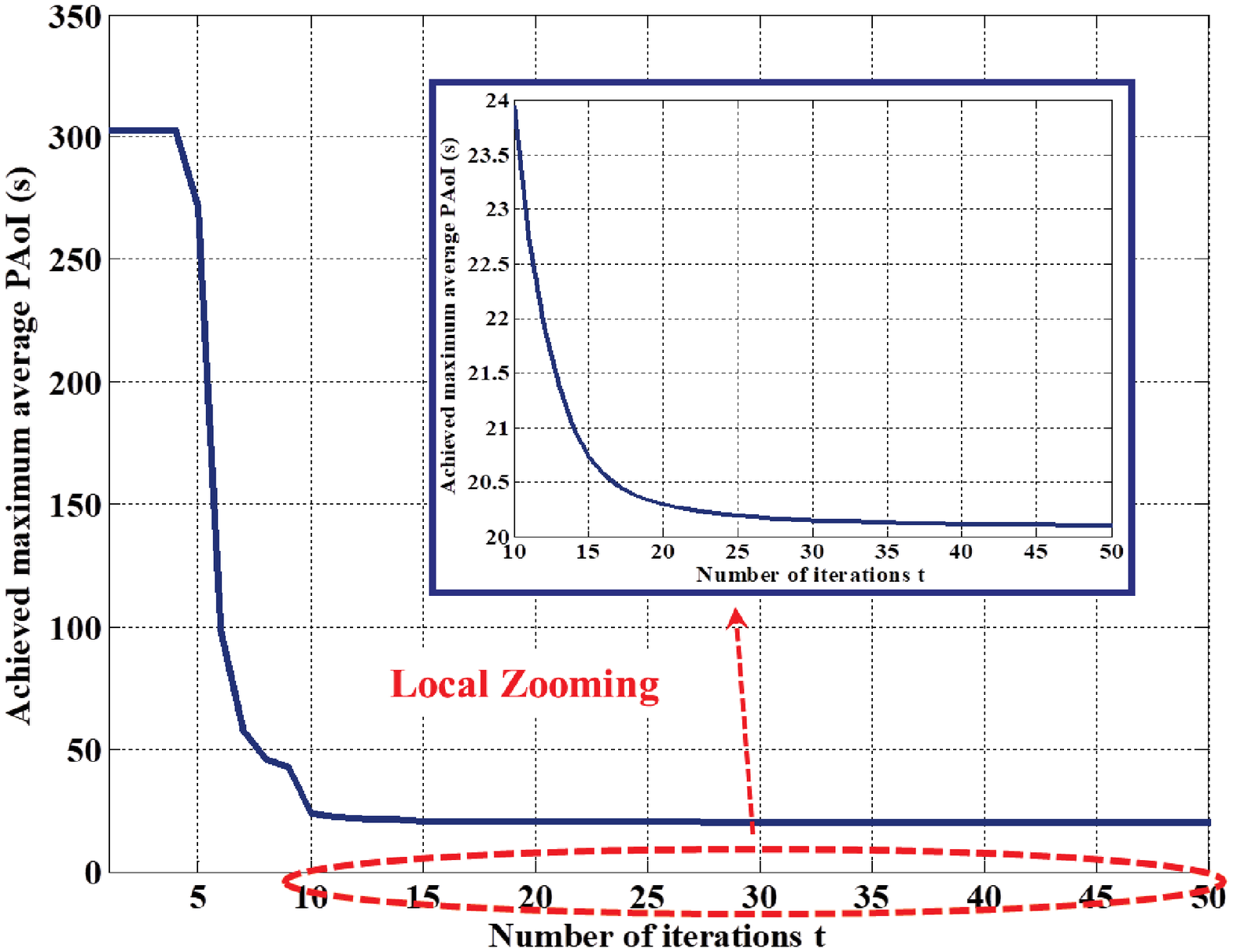}}%
\subfigure[]{\centering \includegraphics[width=2.35in]{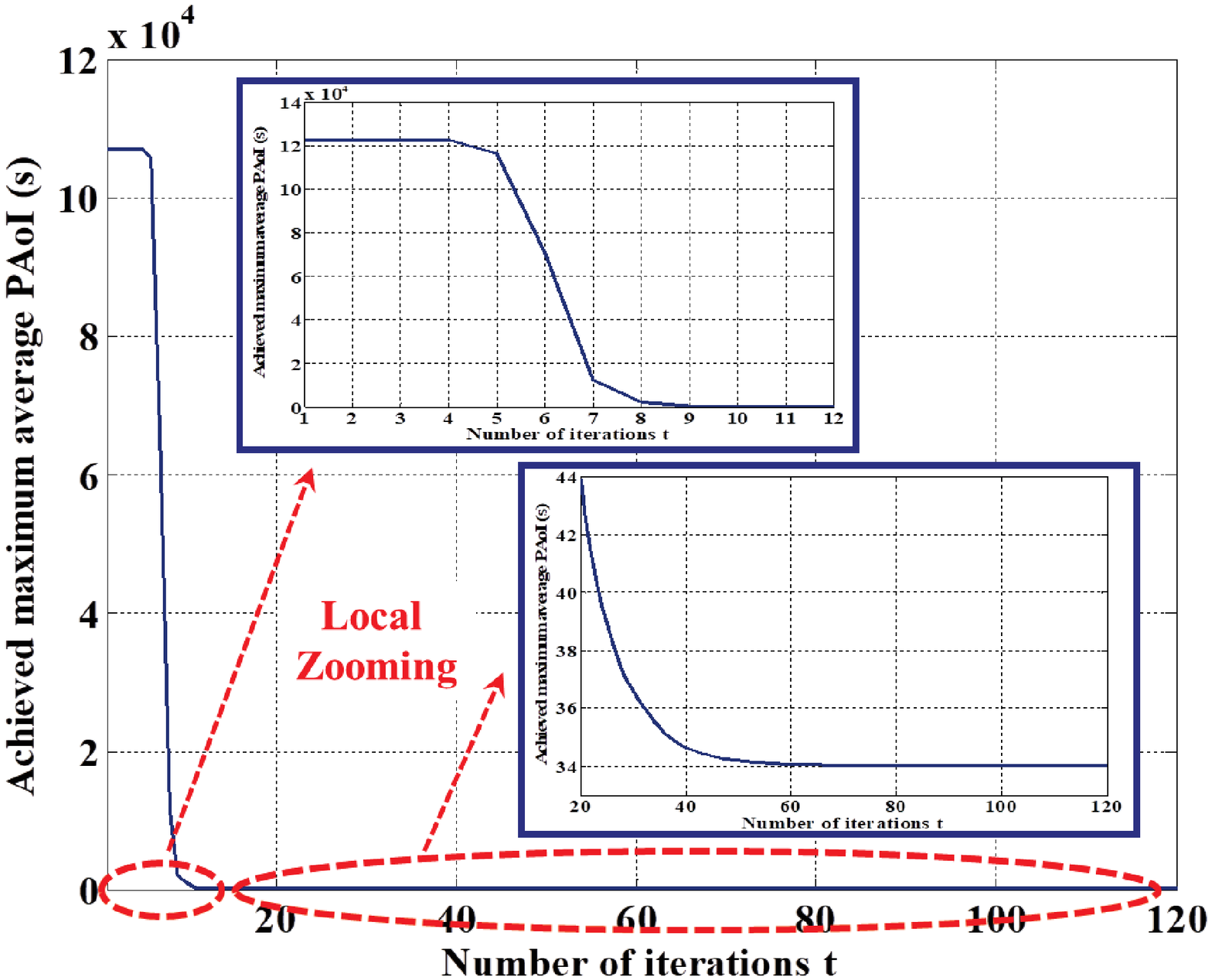}}%
\subfigure[]{\centering \includegraphics[width=2.35in]{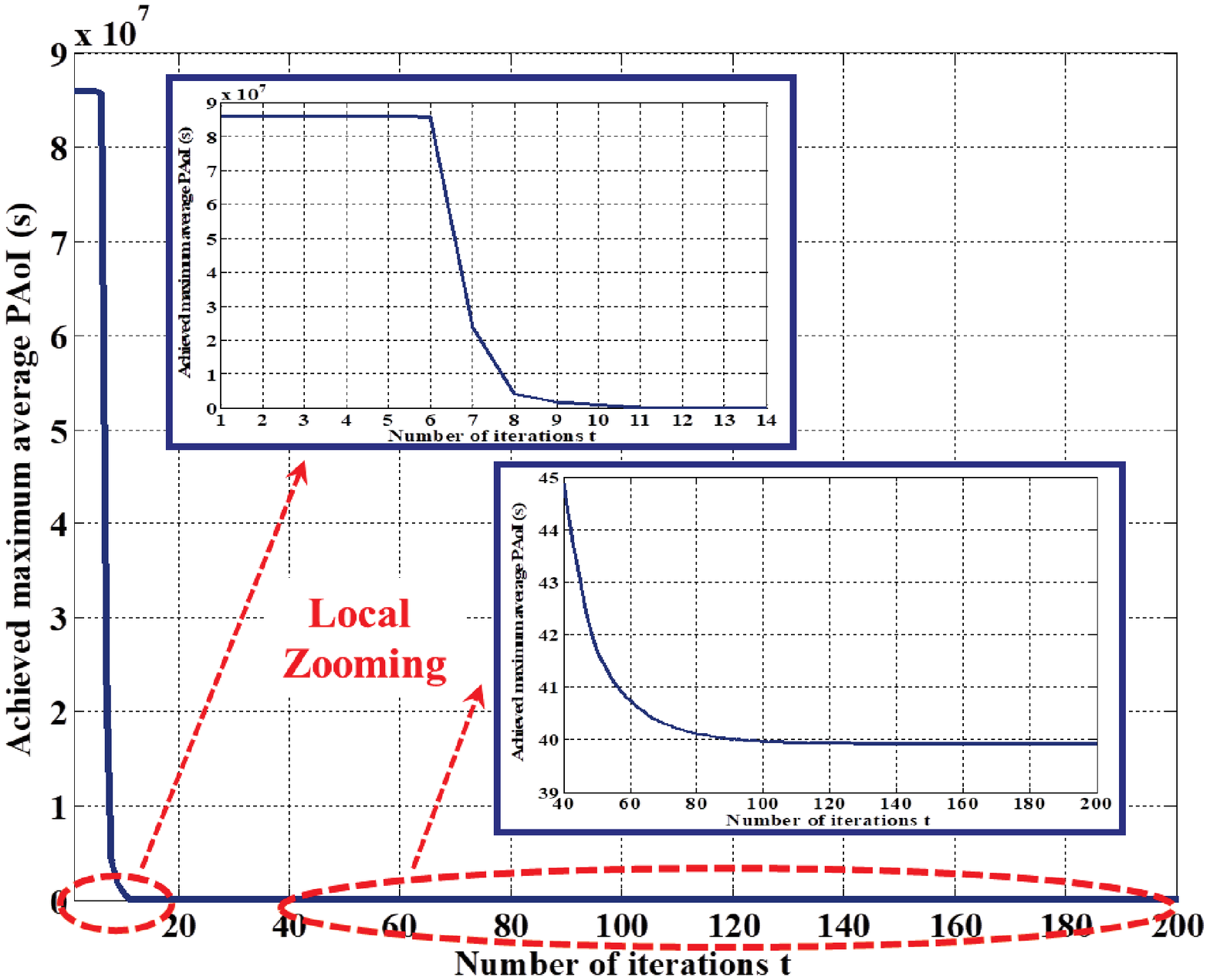}}%
\centering
\centering \caption{Convergence of our proposed algorithm GAP when the number of sensors is set as: (a) $J=2$; (b) $J=6$; (c) $J=10$, while the size of processed data packets is identical and set to 4 Mbits.} \label{Fig:Convergency_GAP_4Mb}
\end{figure*}

It should be noted that, as suggested in \cite{GSS_Linearly_Constraint_Opt}, besides the directions shown in $\mathcal S_{\mathcal T( {{\bf{\Lambda } },\varepsilon } )} $ the generators of $\mathcal N( {{\bf{\Lambda } },\varepsilon } )$ should also be chosen as searching directions by utilizing appropriate step-sizes, which may accelerate the convergence especially when the optimal solution is near the boundary of the feasible region. However, for our concerned problem, we can see that if the feasible point is close to the boundary of $\Omega$, then the system approaches the unstable state, i.e., the average PAoI becomes extremely large. In other words, the optimal solution of problem \textbf{P1} is surely an interior point and generally ``far" away from the boundary of $\Omega$. Hence, to avoid unnecessary calculations and accelerate the convergence of the algorithm, we only adopt vectors belonging to $\mathcal S_{\mathcal T( {{\bf{\Lambda } },\varepsilon } )} $ as the candidate searching directions.

After that, we evaluate all obtained trial points. The trial points yielding the function value less than $\widetilde A ( {\bf{\Lambda }} ) - {{10}^{ - 4}}{\varepsilon ^2}$ are regarded as potential starting points for the next iteration. Among them, that yielding the smallest $\widetilde{A}$ is chosen as the best solution after the current iteration to accelerate the convergence, while the potential step-size is kept the same, i.e., line \ref{Alg1:Choosing_Opt_Point}. If there is no such a potential point being found, $\bf{\Lambda }$ will be still adopted as the starting point for the next iteration, while the potential step-size is halved, i.e., line \ref{Alg1:No_Update_Opt_Point}. When Algorithm \ref{Alg:GAP_algorithm} is terminated, one feasible point $\mathbf{\Lambda }^*$ is finally outputted.

The global convergence of our proposed GAP algorithm is guaranteed according to Theorem \ref{Th:Global_Convergency}.
\begin{theorem} \label{Th:Global_Convergency}
For our proposed GAP algorithm, when given a $\Phi _{\min}$ and arbitrary initial guess ${\mathbf{\Lambda } (0)}$, there always exists a positive constant $t_0$ when $t > t_0$ the condition ${\Phi (t)} < {\Phi _{\min}}$ can be satisfied. In other words, the global convergence of this algorithm can be guaranteed.
\end{theorem}

\begin{IEEEproof}
To prove this theorem, we refer to Theorem 5.1 given in \cite{GSS_Linearly_Constraint_Opt}, which presents the sufficient conditions for the global convergence of a generating set search based algorithm. Actually, it could be readily proved that all such requirements on the adopted searching directions, step-sizes and forcing function (i.e., conditions 1, 2, 4, 5 and 6 specified in \cite{GSS_Linearly_Constraint_Opt}) are satisfied by our developed algorithm GAP. Hence, the global convergence of our proposed GAP algorithm can be guaranteed. Readers are referred to the proof in \cite{GSS_Linearly_Constraint_Opt} for detail, which are omitted here due to space limitations.
\end{IEEEproof}

It should be noted that, according to Theorem 6.5 \cite{GSS_Linearly_Constraint_Opt}, if more stringent requirements on the Lipschitz continuity for the gradient of $\widetilde A \left( {\bf{\Lambda }} \right)$ can be satisfied, then our proposed algorithm can globally converge to a local optimal point. However, it is extremely hard to mathematically prove whether such a condition can be met or not, due to the complicated expressions of average PAoI for packets. Even though, the effectiveness of our proposed algorithm could be validated with numerical results in the following subsection.

\subsection{Numerical Results and Discussions}
We now conduct simulations to evaluate the performance of our proposed algorithm. In particular, we vary the number of sensors $J$ from 1 to 10. Meanwhile, the original size of update packets from sensor $S_j$ is set as $24-(j-1)*2$ Mbits, $\forall j \in \mathcal J$, e.g., $C_6 = 24-(6-1)*2 = 14$ Mbits. The equivalent processing rate is the same, 5 Mbits/s, for all update packets, i.e., $\frac {r}{\tau _j} = 5$, $\forall j \in \mathcal J$, and the parameter ${\Phi _{\min}}$ is set to $10^{-5}$.
Furthermore, we set all the processed data packets to the same size, i.e., $\tilde C_j = \tilde C$, $\forall j \in \mathcal J$, which varies in distinct simulation scenarios. The other parameters are given in Table~\ref{Tab:Parameters}. Here, all the simulation results are obtained by averaging over $10^3$ independent runs, and for each run the the potential step-size for searching is initialized with Eq. (\ref{Eq:Initial_Potential_Step_Size}).

Fig. \ref{Fig:Convergency_GAP_4Mb} illustrates the convergence property for various number of sensors. From this figure, two observations are due:
1) for randomly selected feasible points (i.e., ${\mathbf{\Lambda } (0)}$) the resulted maximum average PAoI $ \widetilde{A}$ is extremely high, i.e, the whole system approaches the instable state, especially when more sensors are incorporated. This is mainly due to the fact that compared with the region of update arrival rates with an acceptable lower $ \widetilde{A}$, the ``undesired'' region is much larger, where the randomly selected ${\mathbf{\Lambda } (0)}$ is likely to lie. Thus, making the arrival rate profile $ {\mathbf {\Lambda}} = (\lambda_1, \lambda_2, \cdots, \lambda_J)$ lie in a ``preferable" region to keep the obtained information fresh is very necessary. 2) the convergence rate of GAP is acceptable and does not exponentially increase with the number of sensors. Particularly, when there are 6 sensors our algorithm converges in about 90 iterations. However, when about 67 percent more sensors (i.e., 10 sensors) are incorporated, about 56 percent more iterations (i.e., 140 iterations) are needed before the convergence is achieved. Besides, we note that during the first a few (about 5) iterations no improvement can be made. This is due to the fact, compared with the size of the feasible region $\Omega$ of Problem \textbf{P1}, the adopted initial potential searching step-size is too large. Hence, before it is reduced to a suitable value, the searching direction makes all constraints be violated, i.e., $\mathcal N( {{\bf{\Lambda }},\varepsilon } ) = \mathbb{R}^J$ and $\mathcal T( {{\bf{\Lambda }},\varepsilon } ) = \left\{ \mathbf 0 \right\}$. In other words, during these iterations, no other feasible trial points could be found and hence, no performance improvement can be made. However, when the size of $\Omega$ is not easy to be evaluated, adopting a larger initial searching step-size is recommended, since it can be quickly (exponentially) decreased down to some proper value (seeing line \ref{Alg1:No_Update_Opt_Point} in Algorithm \ref{Alg:GAP_algorithm}).

\begin{figure*}[!t]
\centering
\subfigure[]{\centering \includegraphics[width=3.0in]{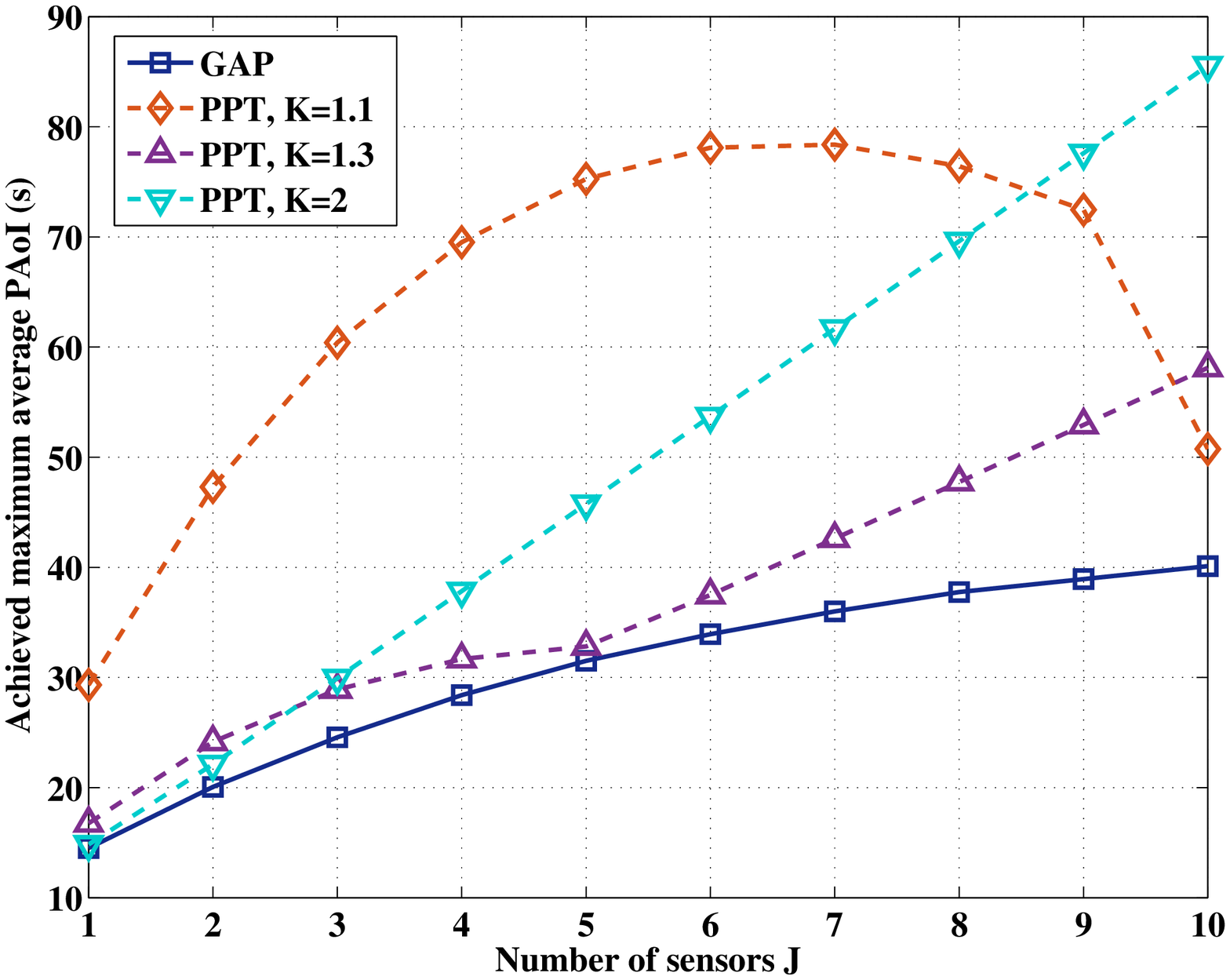}}%
\subfigure[]{\centering \includegraphics[width=3.0in]{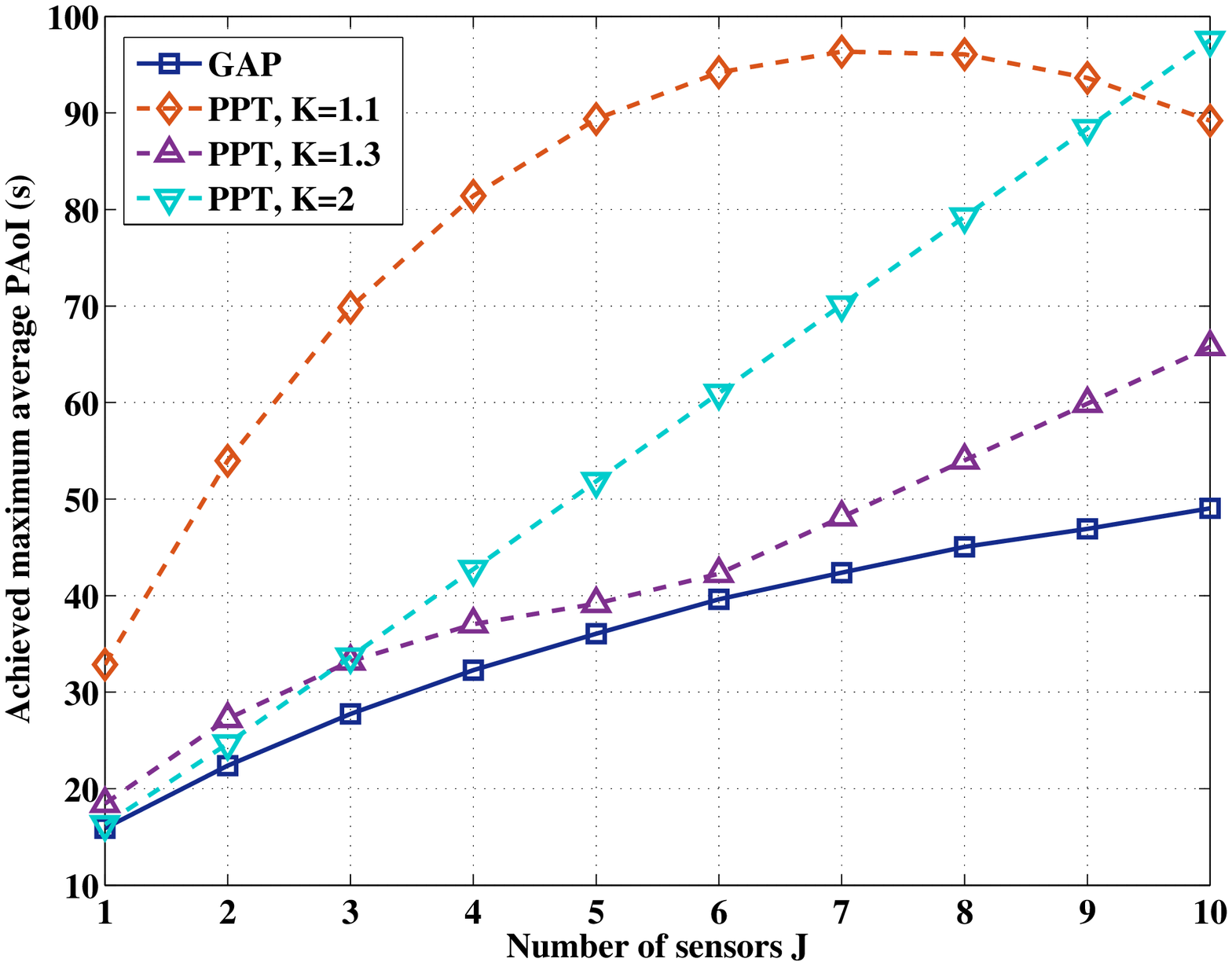}}%
\centering
\centering \caption{Performance comparison in terms of $\widetilde \Delta$ where the size of each processed data packet $\tilde C$ is: (a) 4 Mbits; (b) 1 Mbits.} \label{Fig:Performance_comparison}
\end{figure*}

To evaluate the performance of our proposed algorithm in terms of the achieved maximum average PAoI $\widetilde{A}\left({\mathbf {\Lambda}}\right)$, the performance of a Proportion to the Processing and Transmission rate (PPT) algorithm are considered as the baseline. Concretely, by utilizing the PPT algorithm, the data arrival rates are set as
\begin{align}  \label{Eq:Initial_Guess}
{\lambda _j} = \frac{1}{{KJ}}\min\! \left\{ {\frac{1}{{\mathbb E \big[ {Z_j^P} \big] }},\frac{1}{{\mathbb E \big[ {Z_j^T} \big] }}} \right\},  ~~~ \forall \, \lambda _j \in {\bf{\Lambda }}
\end{align}
where $K > 1$ is a constant affecting the obtained $\widetilde{A}\left({\mathbf {\Lambda}}\right)$. Obviously, the adopted point is feasible, since $\sum_{j = 1}^J {\mathbb E[{Z_j^P}]} {\lambda _j} \leq {1}/{K}$ and $\sum_{j = 1}^J {\mathbb E[ Z_j^T ]} {\lambda _j} \leq {1}/{K}$. In other words, we can control the upper bound of the achieved loads in both processing subsystem and transmission subsystem by adjusting the value of $K$, i.e., a smaller $K$ results in a higher upper bound on loads.

The simulation results are presented in Fig. \ref{Fig:Performance_comparison} (a) and (b), where the size of each processed data packet is set to 4 Mbits and 1 Mbits, respectively. We can observe that our developed GAP algorithm can significantly reduce the achieved maximum average PAoI even when the number of involved sensor is large, although the performance of PPT is occasionally close to that of GAP in some scenarios (e.g., $J = 5$ and $K = 1.3$ in Fig. \ref{Fig:Performance_comparison} (a)). For instance, when there are 10 sensors, the performance improvement is up to $53.13\%$ (reducing from 85.58 s to 40.11 s) and $49.72\%$ (reducing from 97.54s to 49.04s) in Fig \ref{Fig:Performance_comparison} (a) and (b), respectively. The main reason for this improvement lies in the fact that based on the analytical results derived in the previous section, we can essentially capture the joint effect of data preprocessing and transmission on the information freshness, and therefore control the generation rate of updates more efficiently.

\section{Conclusions}

In this paper, we took a fresh look at the problem of optimizing information freshness in computing enabled IoT networks. Considering a system that allows the collected raw data to be preprocessed before transmission, we modeled it as a tandem queue and derived an analytical expression for the average PAoI. Based on the analytical results, we closely examined how computing and transmission affect the information freshness. Furthermore, we developed an algorithm to minimize the achieved maximum average PAoI for updates from different sensors.
Our algorithm is derivative-free and hence applicable to a host of different penalty functions, besides of the maximum of average PAoI. Simulations showed that our algorithm is both efficient and effective, whereas it takes a few steps to converge and largely outperforms the benchmark.

Following our developed framework, several extensions are possible. For instance, when the packets belonging to different sensors are correlated, the framework can be used to develop more advanced processing and optimization schemes. Another future direction is to investigate the scenario where multiple aggregators coexist in the network. Then, an interesting problem is how to align interference and meanwhile maintain information freshness.

\appendices
\section{Proof of Theorem \ref{Theorem:Exp_Waiting_Time_PQ}} \label{Pro:Exp_Waiting_Time_PQ}
\begin{IEEEproof}
We denote the average number of update packets with priority $j$ in the processing queue by $\mathbb E [{N_{j,Q}^P}]$ and the expectation of the remaining processing time of a packet in service by $\mathbb E [{Z_R^P}]$. The expectation $\mathbb E [ {W_1^P} ]$ can be expressed as
\begin{align} \label{Eq:Waiting_Time_S1}
\nonumber \mathbb E \big[ {W_1^P} \big] &={P_{{A_P},B}} \mathbb E \big[ {Z_R^P} \big] +\left(\! {1{ - }{P_{{A_P},B}}} \!\right) \!\cdot 0{\rm{ + }} \mathbb E \big[ {N_{1,Q}^P} \big] \mathbb E \big[ {Z_1^P} \big] \\  \nonumber
&={P_{{A_P},B}} \mathbb E \big[ {Z_R^P} \big] + \lambda_1\mathbb E \big[ {W_1^P} \big] \mathbb E \big[ {Z_1^P} \big] \\
&=\frac{{{P_{{A_P},B}}E \big[ {Z_R^P} \big] }}{{1{\rm{ - }}{\lambda _1}E \big[ {Z_1^P} \big] }}{\rm{ = }}\frac{{{P_{{A_P},B}}E \big[ {Z_R^P} \big] }}{{1{\rm{ - }}{\rho _1}}}
\end{align}
where ${P_{{A_P},B}}$ is expressed in Eq. (\ref{Eq:Busy_Probability_AP}) and ${\rho _1}$ denotes the load in the processing subsystem caused by packets with priority 1. Similarly, for $j>1$ we have
\begin{align} \label{Eq:Waiting_Time_SJ}
\nonumber \mathbb E \big[ {W_j^P} \big] {\rm{ = }}&{P_{{A_P},B}} \mathbb E \big[ {Z_R^P} \big] {\rm{ + }}\left( {1{\rm{ - }}{P_{{A_P},B}}} \right) \cdot 0{\rm{ + }} \\ \nonumber
&\sum_{i = 1}^j {\mathbb E \big[ {N_{i,Q}^P} \big] \mathbb E \big[ {Z_i^P} \big] }+\sum\limits_{i = 1}^{j - 1} {{\lambda _i}\mathbb E \big[ {W_j^P} \big] \mathbb E \big[  {Z_i^P}  \big] }\\
\nonumber {\rm{ = }}&{P_{{A_P},B}}\mathbb E \big[  {Z_R^P} \big]{\rm{ + }}\sum\limits_{i = 1}^j {{\lambda _i}\mathbb E \big[  {W_i^P} \big]\mathbb E\big[  {Z_i^P} \big]}  + \\ \nonumber
& \mathbb E\big[  {W_j^P} \big]\sum\limits_{i = 1}^{j - 1} {{\lambda _i}\mathbb E\big[  {Z_i^P} \big]} \\
{\rm{ = }}&\frac{{{P_{{A_P},B}}\mathbb E \big[ {Z_R^P} \big]  + \sum_{i = 1}^{j - 1} {{\rho _i}\mathbb E \big[ {W_i^P} \big] } }}{{1 - \sum_{i = 1}^j {{\rho _i}} }}.
\end{align}
Directly substituting Eq. (\ref{Eq:Waiting_Time_S1}) into Eq. (\ref{Eq:Waiting_Time_SJ}) and setting $j=2$, we have
\begin{align} \label{Eq:Waiting_Time_S2}
\nonumber \mathbb E \big[ {W_2^P} \big] = &\frac{{{P_{{A_P},B}}\mathbb E \big[ {Z_R^P} \big] + {\rho _1}\frac{{{P_{{A_P},B}}\mathbb E [ {Z_R^P} ] }}{{1{\rm{ - }}{\rho _1}}}}}{{1 - \sum_{i = 1}^2 {{\rho _i}} }} \\
{\rm{ = }}& \frac{{{P_{{A_P},B}}\mathbb E \big[ {Z_R^P} \big] }}{{\left( {1 - \sum_{i = 1}^2 {{\rho _i}} } \right)\left( {1{\rm{ - }}{\rho _1}} \right)}}.
\end{align}
Similarly, for $\forall j >2$, we have
\begin{align} \label{Eq:Waiting_Time_Recurrence}
\nonumber & \left( {1 - \sum_{i = 1}^j {{\rho _i}} } \right)\mathbb E \big[ {W_j^P} \big] = {P_{{A_P},B}}\mathbb E \big[ {Z_R^P} \big] + \sum_{i = 1}^{j - 1} {{\rho _i}\mathbb E \big[ {W_i^P} \big] }  \\ \nonumber
&={P_{{A_P},B}}\mathbb E \big[ {Z_R^P} \big] + \sum_{i = 1}^{j - 2} {{\rho _i}\mathbb E \big[ {W_i^P} \big] }  + {\rho_{j - 1}}\mathbb E \big[ {W_{j - 1}^P} \big]  \\ \nonumber
&\mathop  = \limits^{\left( a \right)}\left( {1 - \sum\limits_{i = 1}^{j - 1} {{\rho _i}} } \right)\mathbb E \big[ {W_{j - 1}^P} \big]  + {\rho_{j - 1}}\mathbb E \big[ {W_{j - 1}^P} \big]  \\
&= \left( {1 - \sum\limits_{i = 1}^{j - 2} {{\rho _i}} } \right)\mathbb E \big[ {W_{j - 1}^P} \big].
\end{align}
where $(a)$ follows Eq. (\ref{Eq:Waiting_Time_SJ}).
Substituting (\ref{Eq:Waiting_Time_S2}) into the recursion formula in (\ref{Eq:Waiting_Time_Recurrence}), we can express the expectation $\mathbb E\big[ {W_j^P} \big]$ as
\begin{align} \label{Eq:Processing_Waiting_Time_Fin}
\mathbb E \big[ {W_j^P} \big] =
\begin{cases}
\frac{{{P_{{A_P},B}}\mathbb E [ {Z_R^P} ] }}{{1{\rm{ - }}{\rho _1}}}, &j=1 \\
\frac{{{P_{{A_P},B}}\mathbb E [ {Z_R^P} ] }}{{\left( {1 - \sum\nolimits_{j = 1}^{J} {{\rho _i}} } \right)\left( {1 - \sum\nolimits_{j = 1}^{J-1} {{\rho _i}} } \right)}}, & j >1
\end{cases}
\end{align}
where $\mathbb E\big[{Z_R^P} \big]$ is the expectation of the remaining processing time of a packet in service.
By applying the renewal-reward theory \cite{Queueing_Performance}, we have
\begin{align} \label{Eq:Remaining Processing_Time_Org}
\mathbb E \big[ {Z_R^P} \big] = \frac{{\mathbb E \Big[ \left( {{Z^P}} \right)^2 \Big] }}{{2\mathbb E \big[ {{Z^P}} \big] }}
\end{align}
where
\begin{align} \label{Eq:Processing_Time_1M}
\mathbb E \big[ {{Z^P}} \big] = \sum_{j = 1}^J {  {\frac{{{\lambda _j}}}{{\sum_{i = 1}^J {{\lambda _i}} }}\mathbb E \big[ {Z_j^P} \big] } }  = \frac{{\sum_{j = 1}^J {{\rho _j}} }}{{\sum_{j = 1}^J {{\lambda _i}} }}
\end{align}
and
\begin{align} \label{Eq:Processing_Time_2M}
\mathbb E\Big[ {{{\left(\! {{Z^P}}\! \right)}^2}} \Big] = \sum_{j = 1}^J {  {\frac{{{\lambda _j}}}{{\sum_{i = 1}^J {{\lambda _i}} }}{{\left( {\mathbb E \big[ {Z_j^P} \big] } \right)}^2}}  }  = \frac{{\sum_{j = 1}^J {{{{{ {{\rho^2_j}} }}}}/{{{\lambda _j}}}} }}{{\sum_{j = 1}^J {{\lambda _i}} }}.
\end{align}

Finally, combining from Eq. (\ref{Eq:Processing_Waiting_Time_Fin}) to (\ref{Eq:Processing_Time_2M}) and introducing the indicator function $\chi_{ \{ \cdot\} }$, we can draw the conclusion shown in Theorem
\ref{Theorem:Exp_Waiting_Time_PQ}.
\end{IEEEproof}

\section{Proof of Theorem \ref{Theorem:Exp_Transmission_Time_TQ}} \label{Pro:Exp_Transmission_Time_TQ}
\begin{IEEEproof}
We note that the expected transmission time varies for data packets from different sensors. According to Eq. (\ref{Eq:Rate_Random_Variable}), we have the transmission time of one data packet from the aggregator to destination node given by
\begin{align} \label{Eq:Time_Random_Variable}
Z_{\tilde{C}}^T = \frac{{{\tilde{C}}}}{{R_D}} = \frac{{{\tilde{C}}\ln 2}}{{B}}\frac{1}{{\ln \left( {1 + \frac{{{p_A}h{d}^{ - {\alpha}}}}{{{\sigma ^2}}}} \right)}}
\end{align}
where $\tilde{C} \in \{\tilde{C}_1, \tilde{C}_2, \cdots, \tilde{C}_J\}$ denotes the size of the concerned packet. Note that $Z_{\tilde{C}}^T$ is a random variable due to the random channel gain. Moreover, $Z_{\tilde{C}}^T$ monotonically decreases with respect to the channel gain $h$
with the expression given by
\begin{align}
h = \frac{{{\sigma ^2}( {{\rm{exp}}( {\frac{{{\tilde{C}}\ln 2}}{{B}}\frac{1}{{Z_{\tilde{C}}^T}}} ) - 1} )}}{{{p_A}{d}^{ - {\alpha}}}} = f( {Z_{\tilde{C}}^T} )
\end{align}
where $f( {Z_{\tilde{C}}^T} )$ is the function inversely mapping from $Z_{\tilde{C}}^T$ to $h$. In consequence, we obtain the cumulative distribution function (CDF) and probability density function (PDF) of $Z_{\tilde{C}}^T$, respectively, as follows
\begin{align} \label{Eq:Time_CDF}
{{\rm{F}}_{Z_{\tilde{C}}^T}}\left( t \right) &{\rm{  =}}P\left( {Z_{\tilde{C}}^T \le t} \right) = \int_{f\left( t \right)}^\infty  {\exp \left( { - x} \right)dx}
\\ \nonumber
&= \exp \left( \frac{{{\sigma ^2}\left( {1 - {\rm{exp}}\left( {\frac{{\tilde C\ln 2}}{B}\frac{1}{{t}}} \right)} \right)}}{{{p_A}{{ {d}}^{ - {\alpha}}}}} \right)
\end{align}
and
\begin{align} \label{Eq:Time_PDF}
{f_{Z_{\tilde{C}}^T}}\left( t \right) &= - \exp \left( { - f\left( t \right)} \right)\frac {df\left( t \right)}{dt} \\ \nonumber
&= \frac{{\tilde C\ln 2{\sigma ^2}}}{{B{p_A}{{ {d}}^{ - {\alpha}}}}}\frac{{\exp \left( {\frac{{\tilde C\ln 2}}{{Bt}} + \frac{{{\sigma ^2}\left( {{\rm{1}} - {\rm{exp}}\left( {\frac{{\tilde C\ln 2}}{{Bt}}} \right)} \right)}}{{{p_A}{{ {d} }^{ - {\alpha}}}}}} \right)}}{{{t^2}}}.
\end{align}

As such, for packets originally generated from sensor $S_j$, the expectation of transmission time can be attained as
\begin{align} \label{Eq:Transsmission_Time_Exp_Sj_Def}
\mathbb E\big[ {Z_j^T} \big] = \mathbb E\Big[ {Z_{\tilde C}^T\left| {\tilde C{\rm{ = }}{{\tilde C}_j}} \right.} \Big] = \int_0^\infty  {t{f_{Z_{\tilde C}^T\left| {\tilde C{\rm{ = }}{{\tilde C}_j}} \right.}}\left( t \right)} dt.
\end{align}
Finally, by substituting (\ref{Eq:Time_PDF}) into (\ref{Eq:Transsmission_Time_Exp_Sj_Def}) we can draw the conclusion in Theorem \ref{Theorem:Exp_Transmission_Time_TQ}.
\end{IEEEproof}

\section{Proof of Theorem \ref{Theorem:Exp_Waiting_Time_TQ}} \label{Pro:Exp_Waiting_Time_TQ}
\begin{IEEEproof}
We adopt the principle of maximum entropy (PME) to derive an approximation of the expectation $\mathbb E\big[ {W_j^T} \big]$, $\forall j \in \mathcal J$. The interested readers are referred to \cite{EM_PL,EM_Journal,EM_Q_1,EM_Q_3} for more details about PME and its applications for performance analysis in various types of queueing systems.

In the transmission queue, the expectation of the waiting time for a typical data packet in the queue can be expressed
\begin{align}   \label{Eq:Waiiting_Time_Transmission_Q_No_Diff_Org}
\mathbb E\big[ {{W^T}} \big] = \sum_{j = 1}^J {P_j^T \mathbb E\big[ {W_j^T} \big]} \mathop  = \limits^{\left( a \right)} \sum\limits_{j = 1}^J {\frac{{{\lambda _j}}}{{\sum_{i = 1}^J {{\lambda _i}}}}\mathbb E\big[ {W_j^T} \big]}
\end{align}
where $P_j^T$ denotes the probability that there is one packet arriving at the transmission queue originally from sensor $S_j$, and $\mathbb E\big[ {W_j^T} \big]$ is the expectation of its waiting time. In addition, (a) holds under the condition that the previous processing subsystem is stable, i.e., the arrivals are all processed on average. Moreover, from Theorem \ref{Theorem:Exp_Transmission_Time_TQ} we have that for a typical packet, the average time spent in the transmission subsystem can be expressed as
\begin{align} \label{Eq:Transsmission_Time_Exp}
\mathbb E\big[ {{Z^T}}\big]&= \sum\limits_{j = 1}^J {\mathbb E\Big[ {Z_{\tilde C}^T\left| {\tilde C{\rm{ = }}{{\tilde C}_j}} \right.} \Big]P\left( {\tilde C{\rm{ = }}{{\tilde C}_j}} \right)} \\ \nonumber
&\mathop  = \limits^{\left( a \right)} \sum_{j = 1}^J {\frac{{{\lambda _j}}}{{\sum_{i = 1}^J {{\lambda _i}} }}\mathbb E\Big[ {Z_{\tilde C}^T\left| {\tilde C{\rm{ = }}{{\tilde C}_j}} \right.} \Big]}  = \frac{{\sum_{j = 1}^J {{\lambda _j}\mathbb E\big[ {Z_j^T} \big]} }}{{\sum_{j = 1}^J {{\lambda _j}} }}
\end{align}
where ($a$) holds under the condition that the previous processing subsystem is stable, and the expectation $\mathbb E\big[ {Z_j^T} \big]$ is given in (\ref{Eq:Transsmission_Time_Exp_Sj}). Then, by applying Little's law to the transmission subsystem and combining the result with (\ref{Eq:Waiiting_Time_Transmission_Q_No_Diff_Org}) we have
\begin{align}   \label{Eq:Waiiting_Time_Transmission_Q_Org}
\mathbb E\big[ {W^T} \big] &= \frac {\mathbb E \big[N^T\big]}{\sum_{j = 1}^J {{\lambda _j}}} - \mathbb E\big[ {{Z^T}} \big] \\ \nonumber
&= \sum\limits_{j = 1}^J {\frac{{{\lambda _j}}}{{\sum_{i = 1}^J {{\lambda _i}} }}\mathbb E\big[ {W_j^T} \big]} = \sum\limits_{j = 1}^J {\frac{{{\lambda _j}{\mu _j}}}{{\sum_{i = 1}^J {{\lambda _i}} }}\mathbb E\big[W_1^T\big]}
\end{align}
where $\mathbb E \big[N^T\big]$ denotes the expectation of the total number of packets in the transmission subsystem, $\mathbb E\big[ {{Z^T}} \big]$ is given by (\ref{Eq:Transsmission_Time_Exp}), and $\mu _j$ represents the ratio $\frac{\mathbb E\big[ {W_j^T} \big]}{\mathbb E\big[ {W_1^T}\big]}, \forall j \in \{1, 2, \cdots, J\}$. According to Eq. (\ref{Eq:Waiiting_Time_Transmission_Q_Org}), we can obtain $\mathbb E\big[ {W_j^T} \big]$ if $\mathbb E \big[N^T\big]$ and $\mu _j, \forall j \in \{2, \cdots, J\}$, are derived.

As $N^T$ is an integer-value random variable, we use the PME to express its probability mass function as follows
\begin{align}   \label{Eq:PMF_Our}
P\left( {{N^T = n}} \right) &= \frac{1}{G}\exp \left( { - \sum\nolimits_{m = 1}^M {{\beta _m}{{\left( n \right)}^m}}} \right) \\ \nonumber
& \mathop  \approx \limits^{(a)} \frac{1}{G}\exp \left( { - {\beta _1 }n} \right), \forall n \in \{0, 1, 2, \cdots\}
\end{align}
where
\begin{align}   \label{Eq:PMF_Parameter_Z_Our}
G &= \sum\nolimits_{n = 0}^\infty  {\left( { - \sum\nolimits_{m = 1}^M {{\beta _m}{{\left( n \right)}^m}} } \right)}  \\ \nonumber
& \mathop  \approx \limits^{(b)} \sum\nolimits_{n = 0}^\infty  {\exp \left( { - {\beta _1}n} \right)}  = {\left( {1 - \exp \left( { - {\beta _1}} \right)} \right)^{ - 1}}.
\end{align}
Wherein, $\beta_m$ is the introduced Lagrangian multiplier associated with the $m$-th moment of the random variable $N^T$, while (a) and (b) hold due to the first moment approximation\footnote{Note that the accuracy of the approximation improves when more moments of $N^T$ are incorporated, giving rise to a higher complexity. }.

By applying Little's law to the transmission queue and combining the result with (\ref{Eq:PMF_Our}) and (\ref{Eq:PMF_Parameter_Z_Our}) we have the following
\begin{align}   \label{Eq:PMF_Zero}
P\left( {{N^T = 0}} \right)  = 1-\rho ^T  \approx  \left( {1 - \exp \left( { - {\beta _1}} \right)} \right)
\end{align}
where $\rho ^T=\sum_{j = 1}^J {{\lambda _j}} \mathbb E\big[ {{Z_j^T}} \big]$ denotes the probability that the server is busy. As such, using the PME for another time, we have
\begin{align}   \label{Eq:Transmission_Q_User_Num}
{\mathbb E \big[N^T\big]} & \thickapprox \frac{{\partial \ln \left( {1 \!-\! \exp \left( { \! - {\beta _1}} \! \right)} \right)}}{{\partial {\beta _1}}} = \frac{{\sum_{j = 1}^J {{\lambda _j}} \mathbb E\big[ {{Z_j^T}} \big]}}{{1 - \sum_{j = 1}^J {{\lambda _j}} \mathbb E\big[ {{Z_j^T}}\big]}}.
\end{align}

Next, we analyze the ratio $\mu _j$. In the transmission queue, the waiting time of one arriving data packet is related to the number and kinds of packets (i.e., the profile of packets) waiting in front of it, which are determined by the output of the previous processing queue and are extremely difficult to obtain. This is due to the fact that, as previously stated, deriving the PDF of the inter-departure time of packets in the previous priority M/G/1 queue is hindered due to the complexity. Next, we derive an approximation of ratio $\mu _j, \forall j \in \{2,3,\cdots, J\}$ by recalling the analysis for $\mathbb E\big[W_j^P\big]$ in Appendix \ref{Pro:Exp_Waiting_Time_PQ}. Particularly, for one typical packet originally generated by sensor $Sj$, during its waiting time in the processing queue, the profile of packets which are severed before it can be statistically expressed as
\begin{align}   \label{Eq:Profile_Packet}
\mathbf H_j^P=(H_{j,1}^P, H_{j,2}^P, \cdots, H_{j,j-1}^P, H_{j,j}^P), \forall j \in \mathcal J
\end{align}
where $H_{j,i}^P={\lambda _i}\left( {\mathbb E \big[ {W_i^P} \big] + \mathbb E \big[ {W_j^P} \big] } \right), \forall i \in \{1, 2, \cdots, j-1\}$ and $H_{j,j}^P = {\lambda _j}\mathbb E \big[ {W_j^P} \big]$. We note that when this typical packet is waiting or being served in the processing queue, packets in front of it would be sequently sent into the transmission queue and served by the transmitter. Hence, recalling the analysis for $\mathbb E\big[W_j^P\big]$ in Appendix \ref{Pro:Exp_Waiting_Time_PQ}, we can derive an approximation of ratio $\mu _j, \forall j \in \{2,3,\cdots, J\}$ as shown in (\ref{Eq:Waiiting_Time_Transmission_Q_Ratio}). It intuitively means that for a typical packet its waiting time spent in the transmission queue is approximately proportional to the difference of its experienced system time (sum of waiting time and service time) in the processing system and the total transmission time of packets described by  Eq. (\ref{Eq:Profile_Packet}).

%

Finally, substituting (\ref{Eq:Transmission_Q_User_Num}) and (\ref{Eq:Waiiting_Time_Transmission_Q_Ratio}) into (\ref{Eq:Waiiting_Time_Transmission_Q_Org}) we can draw the conclusion shown in Theorem \ref{Theorem:Exp_Waiting_Time_TQ}.
\end{IEEEproof}



\bibliographystyle{IEEEtran}
\bibliography{IEEEabrv,CUTD_Ref}
\end{document}